\documentclass[showpacs,10pt,aps,prd,reprint,onecolumn,notitlepage,nofootinbib,superscriptaddress,english]{revtex4-1}
\usepackage{mathptmx}
\usepackage[T1]{fontenc}
\usepackage[a4paper]{geometry}
\usepackage{babel}

\usepackage{amsmath}
\usepackage{graphicx}
\usepackage{amssymb}
\usepackage{esint}
\usepackage[unicode=true, 
 bookmarks=true,bookmarksnumbered=false,bookmarksopen=false,
 breaklinks=false,pdfborder={0 0 1},backref=false,colorlinks=false]
 {hyperref}
\hypersetup{pdftitle={Reconstructing a model of implicit shapes from an unorganized point set},
 pdfauthor={Levente Hunyadi and István Vajk},
 pdfkeywords={model reconstruction, alternating optimization, total least squares, self-organizing method}}
 
\usepackage{hyperref}
\usepackage{amsfonts}
\usepackage{tensor}
\usepackage{graphicx}
\usepackage[]{subfigure}
\usepackage[usenames, dvipsnames]{color}

\hypersetup{
    bookmarks=true,         
    pdftoolbar=true,        
    pdfmenubar=true,        
    pdffitwindow=true,     
    pdfstartview={FitH},     
    pdftitle={My title},     
    pdfauthor={author},      
    pdfsubject={Subject},    
    pdfcreator={Creator},    
    pdfproducer={Producer},  
    pdfkeywords={keyword1} 
    pdfnewwindow=true,       
    colorlinks=true ,        
    linkcolor=blue  ,        
    citecolor=blue,      
    filecolor=green,       
    urlcolor=blue            
}

\providecommand{\ed}{\mathrm{d}}

\begin{document}

\title{Gravitational Collapse in {{Repulsive}} $R+\mu^4/R$ Gravity}

\author{Mohsen Fathi}
\email{m.fathi@shargh.tpnu.ac.ir; \,\,mohsen.fathi@gmail.com}

\author{Morteza Mohseni}
\email{m-mohseni@pnu.ac.ir}

 \affiliation{Department of Physics, 
Payame Noor University (PNU),
P.O. Box 19395-3697 Tehran, Iran}

\begin{abstract}
In this paper we work out collapsing conditions for a spherical star in the weak field limit of the $R+\mu^4/R$ gravity and discuss the importance of the parameter $\mu$ to generate different criteria in the theory. Such criteria are proved to 
be resulting in a variety of different fates for the evolution of the outer shells of stars. Furthermore, we investigate the special case of violating the first junction condition and point out corresponding contradictions to the normal cases. These results show that the consistency of the $R+\mu^4/R$ theory of gravity with the common astrophysical predictions relies highly on the adoption of the parameter $\mu$ and satisfaction/{{violation}} of the first junction condition. For those anomalous results, further observational attempts are mandatory. 

\bigskip

\noindent{\textit{keywords}}: $f(R)$ gravity, Oppenheimer-Snyder collapse,  Generalized Raychaudhuri equation, Spherical stars

\end{abstract}

\maketitle

\section{Introduction}\label{sec:introduction}

To those who believe that the late-time acceleration of the universe can be explained by modified theories of gravity to mimic dark energy, the $f(R)$ gravity theories have always been of great significance during the last decade. Indeed such issues were first discussed in Refs.~\cite{Capozziello2002,Capozziello2003} (also for a more recent study see Ref.~\cite{Revelles2013}). Because of their direct generalizations of the Ricci scalar function in the geometric part of the gravitational action, $f(R)$ theories appear to be the most intuitive modifications to general relativity (for remarkable reviews see Refs.~\cite{Sotiriou2010,DeFelice2010,Capozziello2011}). During the last decade, the $f(R)$ theories have been subjected to serious theoretical investigations in order to check their consistency with observational evidences. For example in Refs.~\cite{Navarro2007,Nojiri2007,Lin2010}, some models have been compared to the local gravitational tests within the solar system. Moreover, the contribution of $f(R)$ gravity in cosmological studies has been advanced by its developers. For example, imposition of observational constraints on several $f(R)$ models in the Palatini formalism \cite{Fay2007}, the relevance of cosmographic parameters to the constraining of higher order $f(R)$ cosmology \cite{Capozziello2008a}, impacts of Noether's symmetry on exact cosmological solutions \cite{Capozziello2008b} and also the emergence of dark energy effects from $f(R)$ theories at large scales \cite{Capozziello2008e} have been discussed. Furthermore, a variety of $f(R)$ extensions to other alternative gravitational theories like Ho\v{r}ava gravity, have been discussed in Ref.~\cite{Chaichian2010} and reviewed in Ref.~\cite{Nojiri2011}, where the authors provide a good platform to unify descriptions of the early-time inflation and late-time universal acceleration. A very peculiar case of the theories has been investigated in Ref.~\cite{Park2011}, where this model was calibrated in such a way that the parameter space close to the $\Lambda$CDM model could be consistent with the observational data. In Ref.~\cite{Cardone2012}, two classes of $f(R)$ theories are set into the cosmic observational results (including the Hubble diagram of type Ia supernovae), and compare the modified gravity predictions to those of dark energy models. It has also been shown that through a negative equation of state, scalar degrees of freedom in power-law and Starobinsky $f(R)$ models, are of considerable importance in producing late-time acceleration (see Ref.~\cite{Goswami2013} and references therein). The above examples are just a sample of the tremendous amount of works which have been devoted to developing $f(R)$ cosmology. However, everything becomes more interesting when the solar system gravitational tests, prompt the necessity of spherically symmetric spacetime geometries extracted from $f(R)$ theories. As it is well-known, such approach is essentially so old and dates back to the famous Schwarzschild spherically symmetric static solution of Einstein field equations. However, it inevitably becomes the most reliable approach to the solar system mass contents as those spherically symmetric objects, responsible for gravitational effects. This approach, as well as in the case of the Schwarzschild solution of the general theory of relativity, constitutes the foundations of the mathematical descriptions of $f(R)$ black holes. An enthusiastic tripper actually is exposed to a rather extensive field of study once he/she turns to fundamental facets of such astrophysical objects. However it makes sense to begin with available spherically symmetric solutions to $f(R)$ field equations to construct a valid framework. So far, such solutions have been obtained and discussed within several classes of $f(R)$ theories (for example see Refs.~\cite{Capozziello2007a,Kainulainen2007,Capozziello2008d,Nzioki2010,Arbuzova2014} for spherically symmetric solutions and Refs. \cite{Sharif2010,Capozziello2010a} for some other features, including axial symmetry). Static solutions of different $f(R)$ models, regarded as generators of the exterior geometries of static $f(R)$ black holes, have also been studied in terms of their general features in the context of Schwarzschild-(anti) de Sitter black holes \cite{Dombriz2009a,Dombriz2010a} and charged spherically symmetric black holes introducing a Maxwell source \cite{Hendi2010a,Aghmohammadi2010a,Hendi2011a,Hendi2012a} (see also Ref.~\cite{Laurentis2012a} for a good review). These studies have been generalized to classes of solutions of spherically symmetric spacetimes for non-linearly charged sources in Refs.~\cite{Rodrigues2015,Rodrigues2016}, where the near horizon effects regarding the satisfaction/violation of the strong energy condition (SEC) have been discussed. Extra dimensional charged $f(R)$ black holes have also been investigated in Ref.~\cite{Amirabi2016}. 

However, facing the concept of black holes, one's mind retrieves the final fate of a super-massive star once it has consumed its nuclear fuel and embarks on shrinking until it reaches a certain gravitational radius. This is what that was firstly highlighted in 1939 by Oppenheimer and Snyder in their outstanding paper "{\it{On Continued Gravitational Contraction}}" \cite{Oppenheimer1939}. There, they treated an observer, comoving with a star's surface with interior and exterior geometries given respectively by the Friedmann-Lama\^{i}tre-Robertson-Walker (FLRW) and Schwarzschild metrics. Once the interior and exterior regions were linked, the observer would detect a fatal contraction of the star's surface. This process is known as the Oppenheimer-Snyder (OS) collapse {{which provides a collapsing geometrical framework known as the OS spacetime}}. The large popularity of studying the OS collapse of spherical stars, as well as in general relativity, has also taken place in the context of alternative theories like $f(R)$ gravity. The general cases of perfect fluid stars have been discussed in Refs.~\cite{Sharif2011,Shamir2015} where it has been suggested that the scalar curvature itself can function as the source of a repulsive force which erodes the collapsing process. Such perfect fluid stars have proven to possess an apparent horizon if some certain initial conditions are satisfied \cite{Chakrabarti2016}. Additionally, the OS collapse in some specific examples of $f(R)$ theories was given rigorous attentions in Refs.~\cite{Goswami2014,Said2014}. Moreover, the spherical OS collapse effects on the scalar degree of freedom $f'(R)$, have been proved to claim the transition from general relativity to $f(R)$ gravity during the reinforcement of the gravitational energy \cite{Guo2014}. 

In this paper as well, we study the OS collapse of a spherical perfect fluid star within a specific class of $f(R)$ theories. Initially, this theory was introduced in the form of $f(R)=R-\mu^{2(n+1)}/{R^n} \,\,\, (n>0)$
\cite{Capozziello2003,Capozziello2003b,Nojiri2003,Carroll2004} with particular attention to the special case of $n=1$ (like in the current study) in Ref.~\cite{Carroll2004}. The model was intended to describe the late-time acceleration, but soon it was revealed that this theory suffers from a variety of problems such as matter instability \cite{Dolgov2003}, lack of a matter domination era \cite{Amendola2007a,Amendola2007b} and inconsistency with the solar system constraints \cite{Navarro2007,Chiba2003,Olmo2005a,Olmo2005b,Erickcek2006,Chiba2007}. These problems mostly originate in the fact that the model suggests that $f''(R)<0$. So to overcome this dilemma, one way is to adjust the theory to $f(R)=R+\mu^{2(n+1)}/{R^n}$. In this way, the consequent condition $f''(R)>0$ can significantly heal the shortages of the original model; the stability is retrieved \cite{Nojiri2003} and a consistent matter domination era is retained \cite{Sawicki2007}. Furthermore, the conformity with the standard local gravity tests in the solar system has been elaborated in Ref.~\cite{Saaidi2010}, where a static spherically symmetric solution for the weak field version of $R+\mu^4/R$ gravity has been used. Generally and in the geometric units ($G=c=1$), the relevant field equations are obtained by imposing $f(R)=R+\mu^4/R$ in the modified action
\begin{equation}\label{eq:action}
\mathcal{S}_{f(R)}=\frac{1}{16\pi}\int{\ed^4x\,\sqrt{-g}\,f(R)}+\int{\ed^4x\, \mathcal{L}_m(g_{\alpha\beta},\Psi_m)},
\end{equation}
where $\mathcal{L}_m$ is the matter Lagrangian which is described in terms of the spacetime metric $g_{\alpha\beta}$ and matter fields $\Psi_m$. The standard
variation with respect to the metric gives the field equations
\begin{equation}\label{eq:FieldEqn}
\left(1-\frac{\mu^4}{R^2}\right) R_{\alpha\beta}-\frac{1}{2} \left(1+\frac{\mu^4}{R^2}\right) R\, g_{\alpha\beta}-\mu^4 \left(\nabla_\alpha\nabla_\beta
-g_{\alpha\beta \,\Box}\right) \frac{1}{R^2} = 8 \pi T_{\alpha\beta},
\end{equation}
where $\Box\equiv\nabla^\alpha\nabla_\alpha$ and $T_{\alpha\beta} = -\frac{2}{\sqrt{-g}} \frac{\delta \mathcal{L}_m}{\delta g^{\alpha\beta}}$ is the energy-momentum tensor. In Ref.~\cite{Saaidi2010}  a spherically symmetric solution to the vacuum field equations of Eq.~(\ref{eq:FieldEqn}) was proposed in the form
\begin{equation}\label{eq:SolutionGeneral}
\ed s^2=-A(r) \ed t^2 + B(r)^{-1} \ed r^2 + r^2 \ed \Omega^2
\end{equation}
with
\begin{subequations}\label{eq:SolutionSpecified}
\begin{align}
A(r) &{} ={} 1 - \frac{2 M}{r} + \frac{3}{4} \kappa (\mu r)^{\frac{4}{3}},\label{eq:SolutionSpecified-A(r)}
\\
B(r) &{} ={} 1 - \frac{2 M}{r} +  \kappa (\mu r)^{\frac{4}{3}},\label{eq:SolutionSpecified-B(r)}
\end{align}
\end{subequations}
in which $\kappa = \left(\frac{4}{147}\right)^{\frac{1}{3}}$, $\mathrm{dim}[M] = \mathrm{m}$ and $\mathrm{dim}[\mu] = \mathrm{m}^{-1}$. The above solution, determines the exterior geometry of a spherically symmetric massive object of mass $M$, and the impacts of the $R+\mu^4/R$ theory is only supposed to generate a perturbation on the Schwarzschild geometry. Therefore this solution is pretty useful in inspecting the behavior of the theory in the solar system. However in Ref.~\cite{Fathi2015} we showed that some certain ranges of $\mu$ within the $R+\mu^4/R$ theories correspond to the existence of trapped surfaces on the Schwarzschild radius $r=2 M$. Such surfaces signal the existence of trapping horizons which are traits of black hole regions caused by the gravitational collapse. These regions are hence, providers of cosmic censorship \cite{Wald1999}. 

In this paper therefore, we study such collapse of the outer shell of a spherical star, beyond which the geometry is described by the line element in Eqs.~(\ref{eq:SolutionGeneral}) and (\ref{eq:SolutionSpecified}). Our aim is the determination of specific ranges for the parameter $\mu$ which can provide {{an attractive gravitational force in order to be consistent with the}} spherical OS collapse for relevant massive objects. Such relevance is proved to be relying on the mass relation of a star, to its density and radius. We show that such relation in the $R+\mu^4/R$ theories is not simply given by the multiplication of density by volume. The mass contents of a typical star then make it possible for adoption of reliable values of $\mu$. Regarding the outer shell of a star as a time-like congruence, we independently investigate the possibility of the collapse of the outer shell, through evaluating the {{attraction/repulsion}} conditions and the SEC. This will be done by applying that time-like congruence in a geodesic Raychaudhuri equation. The results confirm that once both the junction conditions are satisfied, the collapse criteria are {{different from those in general relativity (collapsing within repulsive gravity), but approximately the same as those for ultra massive objects.}} However the crucial role of $\mu$ can be discriminated for heavier objects like neutron stars. In the $R+\mu^4/R$ theory, such objects appear to be more unstable than they are in general relativity. However, violating the first junction condition, the time-like shell is no more geodesic and normalized and we will take care of the situation by means of a generalized version of the Raychaudhuri equation for non-normalized tangential vectors. In this way, we argue that stars who lose the connection between interior and exterior regions, exhibit anomalous properties in the $R+\mu^4/R$ theory of gravity; attractive gravity for stars of masses below the Chandrasekhar limit. For higher masses, once again, diverse values of $\mu$ tell significantly different stories about the {{collapsing criteria (collapse within repulsive gravity and holding up within attractive gravity)}}. Based on their predictions, we can therefore classify the $R+\mu^4/R$ theory of gravity in term of values of $\mu$.

The paper is organized as follows: in Sec.~\ref{sec:massOS} we bring the essentials of the OS collapse of a time-like shell of a spherical star and calculate the mass relation. In Sec.~\ref{sec:convergenceLinked}, we work out the SEC for a time-like star's surface moving on a geodesic and based on different values of $\mu$, we demonstrate the possibility of a congruence convergence and therefore an OS collapse. The same analysis is taken care of in Sec.~\ref{sec:convergenceNon-normalized} for a star which has lost the connection between its interior and exterior geometries. We introduce the generalized Raychaudhuri equation and apply it to determine the extent of disconnection for a specific example. This example is specialized for typical stars of definite masses and collapse conditions are discussed. Concluding remarks are given in Sec.~\ref{sec:conclusion}. Throughout this paper, the Greek indices $\alpha,\beta,\gamma, ...$ are 4-dimensional whereas the Latin ones, i.e. $a, b, c, ...$ are 3-dimensional. {{Furthermore, the positive parts of the square roots are considered.}}

\section{Spherical Oppenheimer-Snyder Collapse of a Spherical Shell and the Total Mass Content}\label{sec:massOS}

Let us assume that the star's surface is described by a 3-dimensional time-like hypersurface $\Sigma$, which separates its interior region $\mathcal V^-$ from its exterior region $\mathcal V^+$. Considering $\ell_{\alpha\beta}(x_-^\alpha)$ and $g_{\alpha\beta}(x_+^\alpha)$ to be respectively the metrics of the interior and exterior spacetime manifolds, then the induced metric (or the first fundamental form), intrinsic to the outer and inner sides of $\Sigma$ (i.e. $\Sigma_-$ and $\Sigma_+$) can be obtained by restricting the line element to displacements on either sides of $\Sigma$. This can be done by introducing the intrinsic coordinates $y^a$ on both sides of $\Sigma$, which satisfy the parametric equation $x_\pm^\alpha = x_\pm^\alpha(y^a)$. This way, we can define three tangential 4-vectors  \cite{Poisson2009}
\begin{equation}\label{eq:TangentialVectors-Genral}
{\epsilon_\pm^\alpha}_a = \frac{\partial x_\pm^\alpha}{\partial y^a}
\end{equation}
on $\Sigma_\pm$. For any 4-vectors $n_\pm^\alpha$ normal to $\Sigma_\pm$, we have therefore $n^\pm_\alpha {\epsilon_\pm^\alpha}_a = 0$, where $n^\pm_\alpha$ is the covector associated to $n_\pm^\alpha$ and here we assume that $n_\pm^\alpha n^\pm_\alpha = 1$. Furthermore, we let $n^\alpha$ point from $\mathcal V^-$ to $\mathcal V^+$.  Having these quantities in hand, the line elements intrinsic to $\Sigma_\pm$ are given by \cite{Poisson2009}
\begin{subequations}\label{eq:IntrinsicMetrics-Genral}
\begin{align}
\gamma^+_{ab} &{}={} g_{\alpha\beta}~ {\epsilon_+^\alpha}_a {\epsilon_+^\beta}_b,\label{eq:IntrinsicMetrics-Genral-+}
\\
\gamma^-_{ab} &{}={} \ell_{\alpha\beta}~ {\epsilon_-^\alpha}_a {\epsilon_-^\beta}_b.\label{eq:IntrinsicMetrics-Genral--}
\end{align}
\end{subequations}
Now if we intend to create a smooth joining between $\mathcal V^-$ and $\mathcal V^+$ at the hypersurface $\Sigma$, then the intrinsic metric has to be the same on both $\Sigma_\pm$. This is called the first junction condition and is denoted by \begin{equation}\label{eq:1stJunction-General}
\left[\gamma_{ab}\right]\equiv\gamma_{ab}(\mathcal{V}^+)|_\Sigma - \gamma_{ab}(\mathcal{V}^-)|_\Sigma = \gamma_{ab}^+|_{\Sigma_+}-\gamma_{ab}^-|_{\Sigma_-} = \mathbf{0}.
\end{equation}
Equation (\ref{eq:1stJunction-General}) provides six independent equations. In general relativity, the first junction condition is applied in order to be assured that $\ell_{\alpha\beta}$ and $g_{\alpha\beta}$ together form a valid solution to Einstein field equations \cite{MTW1973}. It has been argued that these two metrics are not in need of any coordinate transformations in order to satisfy the smooth joining over $\Sigma$ \cite{Bonnor1981}. Moreover, if the enveloping spacetimes contain only positive energies, the first junction condition is automatically satisfied \cite{Marolf2005}. However this cannot be guaranteed in alternative theories like $f(R)$ gravity, because the discussions in Ref.~\cite{Marolf2005} have been largely based on the null energy condition in the context of general relativity. In this paper, we also consider the case of violating the first junction condition and argue that if the interior and  exterior regions lose their smooth connection, at some points the exterior segment of the hypersurface, i.e. $\Sigma_+$, can simply overlook the interior segment and therefore it causes a prompted gravitational collapse {{in attractive gravity,}} even in smaller stars.  \\

To obviate the mentioned smooth transition across $\Sigma$, it is also demanded that the extrinsic curvature be the same on both $\Sigma_\pm$. The extrinsic curvature (or the second fundamental form) of a hypersurface is a symmetric tensorial object which is a useful tool to expresses the embedding of a hypersurface into its enveloping space. If the hypersurface is the one that does a temporal foliation of a time-like congruence, then the extrinsic curvature can be used to accentuate the hypersurface concavity/convexity and the corresponding convergence/divergence of the congruence. Based on the quantities which were pointed out earlier in the section, the extrinsic curvatures of the hypersurfaces $\Sigma_\pm$ are defined as
\begin{equation}\label{eq:extrinsicCurvature-General}
K^\pm_{ab} = n_{\alpha;\beta} ~{\epsilon_\pm^\alpha}_a {\epsilon_\pm^\beta}_b,
\end{equation}
with $K^\pm_{ab} = K^\pm_{ba}$. Accordingly, the second junction condition is written as
\begin{equation}\label{eq:2stJunction-General}
\left[K_{ab}\right]\equiv K_{ab}(\mathcal{V}^+)|_\Sigma - K_{ab}(\mathcal{V}^-)|_\Sigma = K_{ab}^+|_{\Sigma_+}-K_{ab}^-|_{\Sigma_-} = \mathbf{0}.
\end{equation}
If the second junction condition is violated, the hypersuface acquires a stress-energy tensor and spacetime becomes singular at $\Sigma$. The violation of the second junction condition in $f(R)$ gravity has been investigated in Ref.~\cite{Senovilla2013}.\\

\begin{figure}[t]
\center{\includegraphics[width=9 cm]{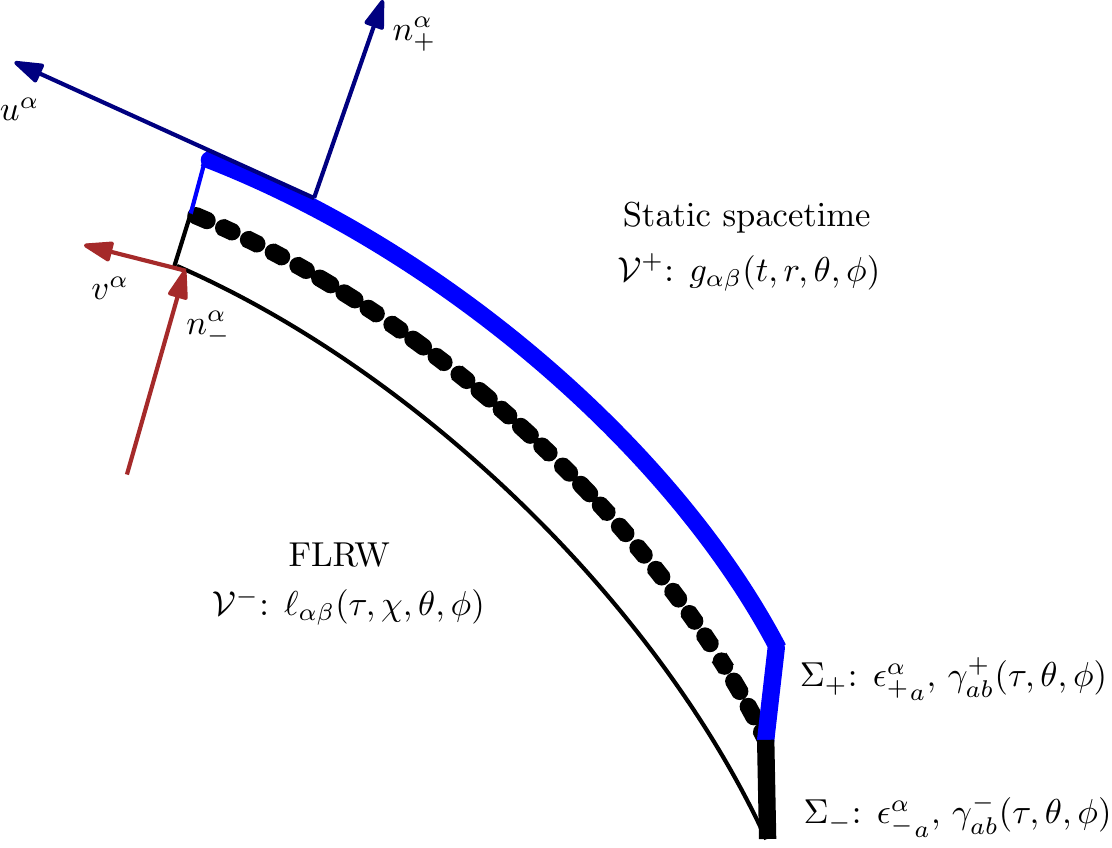}
\caption{\label{fig:Sigma-DoubleLayer} The star's surface $\Sigma$ which partitions the spacetime into its interior and exterior regions. }}
\end{figure}

Now as it has been demonstrated in Fig.~\ref{fig:Sigma-DoubleLayer}, let the region $\mathcal V^-$ be described by the FLRW metric 
\begin{equation}\label{eq:FLRWmetric}
\ed s^2 = -\ed \tau^2 + a(\tau)^2 \left(\ed\chi^2+\sin^2\chi\ed\Omega^2\right),
\end{equation}
which on the comoving world lines (i.e. with $\chi$, $\theta$ and $\phi$ being independent of $\tau$), describes the interior spacetime of a spherical collapsing dust.
In Eq.~(\ref{eq:FLRWmetric}), $\tau$ is the proper time of a comoving fluid and $a(\tau)$ is the scale factor. On the other hand, the region $\mathcal V^+$ is described by the static line element in Eq.~(\ref{eq:SolutionGeneral}). In $\mathcal{V}^+$, the parameter $\tau$ serves as the curve parameter. Therefore the interior and exterior metrics would be $\ell_{\alpha\beta} = \mathrm{diag} \left(-1, a(\tau)^2, a(\tau)^2 \sin^2\chi, a(\tau)^2\sin^2\chi\sin^2\theta\right)$ and $g_{\alpha\beta} = \mathrm{diag}\left(-A(r(\tau)), B(r(\tau))^{-1}, r(\tau)^2, r(\tau)^2\sin^2(\theta(\tau))\right)$. This leads us to adopt a suitable choice for the intrinsic coordinates on both sides of $\Sigma$ as $y^a=(\tau, \theta, \phi)$. Regarding this, using Eq.~(\ref{eq:TangentialVectors-Genral}), the tangential vectors  will be
\begin{subequations}\label{eq:TangentialVectors-Sigma_-}
\begin{align}
{\epsilon_-^\alpha}_1 &\equiv v^\alpha = (1,0,0,0),\label{eq:TangentialVectors-Sigma_-1}
\\
{\epsilon_-^\alpha}_2 &= (0,0,1,0),\label{eq:TangentialVectors-Sigma_-2}
\\
{\epsilon_-^\alpha}_3 &= (0,0,0,1),\label{eq:TangentialVectors-Sigma_-3}
\end{align}
\end{subequations}
on $\Sigma_-$ and 
\begin{subequations}\label{eq:TangentialVectors-Sigma_+}
\begin{align}
{\epsilon_+^\alpha}_1 &\equiv u^\alpha = (\dot t, \dot r, 0, 0),\label{eq:TangentialVectors-Sigma_+1}
\\
{\epsilon_+^\alpha}_2 &= (0,0,1,0),\label{eq:TangentialVectors-Sigma_+2}
\\
{\epsilon_+^\alpha}_3 &= (0,0,0,1).\label{eq:TangentialVectors-Sigma_+3}
\end{align}
\end{subequations}
In Eqs.~(\ref{eq:TangentialVectors-Sigma_-1}) and (\ref{eq:TangentialVectors-Sigma_+1}), $v^\alpha$ and $u^\alpha$ are respectively the tangential velocity 4-vectors of the interior and exterior segments of $\Sigma$ (see Fig.~\ref{fig:Sigma-DoubleLayer}). Furthermore, the dot in Eq.~(\ref{eq:TangentialVectors-Sigma_+1}) denotes $\frac{\partial}{\partial\tau}$. The intrinsic 3-metrics on $\Sigma_\pm$ are hence obtained by means of Eqs.~(\ref{eq:IntrinsicMetrics-Genral}), (\ref{eq:TangentialVectors-Sigma_-}) and (\ref{eq:TangentialVectors-Sigma_+}). We have
\begin{subequations}\label{eq:IntrinsicMetrics - +-}
\begin{align}
\gamma^-_{ab} &= \mathrm{diag}\left(-1, a(\tau)^2 \sin^2\chi_0, a(\tau)^2\sin^2\chi_0\sin^2\theta\right),
\label{eq:IntrinsicMetrics - -}
\\
\gamma^+_{ab} &= \mathrm{diag}\left(-A(r) \dot t^2 + B(r)^{-1} \dot r^2,  r^2, r^2\sin^2\theta\right),\label{eq:IntrinsicMetrics - +}
\end{align}
\end{subequations}
in which $\chi_0$ is where the hypersurface is located. Accordingly, the first junction condition (\ref{eq:1stJunction-General}) implies
\begin{subequations}\label{eq:1stJunction-11,22,33}
\begin{align}
&\left[\gamma_{11}\right] ={} 0 ~~\Longrightarrow~ A(r) \dot t = \sqrt{\frac{A(r)}{B(r)} \left(B(r) + \dot r^2\right)} \doteq \beta(r,\dot r),
\label{eq:1stJunction-11}
\\
&\left[\gamma_{22}, \gamma_{33}\right] = {}0 ~~\Longrightarrow~ r(\tau) = a(\tau) \sin\chi_0 
.\label{eq:1stJunction-22,33}
\end{align}
\end{subequations}
As in Fig.~\ref{fig:Sigma-DoubleLayer}, each side of $\Sigma$ is pierced by a future directed normal vector. On $\Sigma_-$, the conditions $n^-_\alpha n_-^\alpha = 1$ and $v^\alpha n^-_\alpha = 0$ provide
\begin{equation}\label{eq:n_- - FLRW}
n^-_\alpha = \left(0, a(\tau), 0, 0\right).
\end{equation}
The $\tau\tau$ component of the interior extrinsic curvature is obtained by exploiting the definition in Eq.~(\ref{eq:extrinsicCurvature-General}). We have
\begin{eqnarray}\label{eq:K-tautau}
K^-_{\tau\tau} &=& n^-_{\alpha;\beta} ~{\epsilon_-^\alpha}_1 {\epsilon_-^\beta}_1
=  n^-_{\alpha;\beta} ~v^\alpha v^\beta \nonumber\\
&=& \left[\left(n^-_\alpha v^\alpha\right)_{;\beta} - n^-_\alpha v\indices{^\alpha_{;\beta}}\right] v^\beta\nonumber\\
&=& -n^-_\alpha v\indices{^\alpha_{;\beta}} v^\beta\nonumber\\
&=& -n^-_\alpha a_-^\alpha,
\end{eqnarray}
where $a_-^\alpha$ is the 4-acceleration, associated with $v^\alpha$. According to the values in Eqs.~(\ref{eq:TangentialVectors-Sigma_-}), (\ref{eq:IntrinsicMetrics - -}) and (\ref{eq:n_- - FLRW}), we get
\begin{subequations}\label{eq:ExtrinsicCurvature(-)-11,22,33}
\begin{align}
&{K_-^\tau}_\tau ={} 0,
\label{eq:ExtrinsicCurvature(-)-11}
\\
&{K_-^\theta}_\theta = {K_-^\phi}_\phi = a(\tau)^{-1} \cot\chi_0,
\label{eq:ExtrinsicCurvature(-)-22,33}
\end{align}
\end{subequations}
to obtain which, we have applied ${K_-^a}_b = \gamma_-^{ac} K^-_{cb}$ where $\gamma_-^{ac}$ is the inverse of $\gamma^-_{ac}$. On $\Sigma_+$, the conditions $n^+_\alpha n_+^\alpha = 1$ and $u^\alpha n^+_\alpha = 0$ give
\begin{equation}\label{eq:n_+ - Static}
n^+_\alpha = \left(\mp \frac{\dot r}{\sqrt{B(r) \dot t^2 - A(r)^{-1} \dot r}}, 
\pm \frac{\dot t}{\sqrt{B(r) \dot t^2 - A(r)^{-1} \dot r}}, 0, 0\right)
\end{equation}
We adopt the minus sign from the 00 component for $n_+^\alpha$ to be future directed and the positive sign from the 11 component for it to be pointing along the direction of $\Sigma_-\rightarrow\Sigma_+$. Applying the condition in Eq.~(\ref{eq:1stJunction-11}), the desired normal covector is simplified to 
\begin{equation}\label{eq:n_+ - Static - new}
n^+_\alpha = \sqrt{\frac{A(r)}{B(r)}}\left(- \dot r, \dot t, 0, 0\right).
\end{equation}
The extrinsic curvature is obtained, if the same procedure is used on $\Sigma_+$ and its relevant characteristics in Eqs.~(\ref{eq:TangentialVectors-Sigma_+}), (\ref{eq:IntrinsicMetrics - +}) and (\ref{eq:n_+ - Static - new}), giving
\begin{subequations}\label{eq:ExtrinsicCurvature(+)-11,22,33}
\begin{align}
&{K_+^\tau}_\tau ={} -n^+_\alpha a_+^\alpha =  \sqrt{\frac{B(r)}{A(r)}} ~\beta',
\label{eq:ExtrinsicCurvature(+)-11}
\\
&{K_+^\theta}_\theta = {K_+^\phi}_\phi = \sqrt{\frac{B(r)}{A(r)}}~\frac{\beta}{r},
\label{eq:ExtrinsicCurvature(+)-22,33}
\end{align}
\end{subequations}
in which $a_+^\alpha \equiv u\indices{^\alpha_{;\beta}} u^\beta$ is the 4-acceleration associated with $u^\alpha$ and prime stands for $\frac{\partial}{\partial r}$. The second junction condition (\ref{eq:2stJunction-General}) can therefore be imposed, by using values in Eqs.~(\ref{eq:ExtrinsicCurvature(-)-11,22,33}) and (\ref{eq:ExtrinsicCurvature(+)-11,22,33}). We have
\begin{subequations}\label{eq:2ndJunction-11,22,33}
\begin{align}
&\left[{K^\tau}_\tau\right] ={} 0~~\Longrightarrow~~\beta\doteq\tilde E = \mathrm{const.},
\label{eq:2ndJunction-11}
\\
&\left[{K^\theta}_\theta\right] ={} 0~~\Longrightarrow~~\tilde E=\sqrt{\frac{A(r)}{B(r)}}~\cos\chi_0,
\label{eq:2ndJunction-22,33}
\end{align}
\end{subequations}
where in deriving Eq.~(\ref{eq:2ndJunction-22,33}) we have applied Eq.~(\ref{eq:1stJunction-22,33}). The above relations assert that the star's surface (as a time-like congruence) moves on a geodesic, being characterized by the constant of motion $\tilde{E}$ as the energy. As we stated before, this motion is under the effect of the interior region of the star. In the OS collapsing process, this region is supposed to be filled up with a perfect fluid, characterized by the stress-energy tensor
\begin{equation}\label{eq:T_{ab}-General}
T_{\alpha \beta }=\left(\rho(\tau)+p(\tau)\right) v_{\alpha } v_{\beta }+p(\tau) \ell_{\alpha \beta }
\end{equation}
with $\rho(\tau)$ and $p(\tau)$ respectively the fluid's energy density and pressure. For our purpose, we assume that the star is made of dust. Therefore, putting $p=0$ and along with Eqs.~(\ref{eq:FLRWmetric}) and (\ref{eq:TangentialVectors-Sigma_-1}), the stress-energy tensor in $\mathcal V^-$ will be $T_{\alpha\beta}=\mathrm{diag}\left(\rho(\tau), 0, 0, 0\right)$. Substituting in the field equations (\ref{eq:FieldEqn}), we reach the Friedmann equation in the interior region of the star in $R+\mu^4/R$ gravity. We have
\begin{eqnarray}\label{eq:Friedmann-1}
&&\mu ^4 a^4 \left(5 \dot a^4+6 \dot a^2+1\right)+36 \left(\dot a^2+1\right)^4+\mu
 ^4 a^5 \left(\dot a^2+3\right) \ddot a+36 a^3 \left(\dot a^2+1\right) \ddot a^3+108 a^2 \left(\dot a^2+1\right)^2 \ddot a^2\nonumber\\
&& +108 a \left(\dot a^2+1\right)^3 \ddot a+2 \mu ^4 a^6 \left(\ddot a^2-\dddot a \dot a\right)\nonumber\\
&&={}96 \pi  a^2 \rho  \left(a \ddot a+\dot a^2+1\right)^3,
\end{eqnarray}
in which for convenience, we have dropped the parameter $\tau$. Equation (\ref{eq:1stJunction-22,33}) implies $\dot a = \dot r/\sin\chi_0$, whereas Eqs.~(\ref{eq:1stJunction-11}) and (\ref{eq:2ndJunction-11,22,33}), can recover
$\dot r=\sqrt{\cos^2\chi_0-B}$, $\ddot r = -\frac{1}{2} B'$ and $\dddot r = -\frac{1}{2} B'' \dot r$. These relations, together with Eq.~(\ref{eq:Friedmann-1}), give
\begin{equation}\label{eq:Friedmann-2}
\mu ^4 r^4 \left(\frac{4 r^2}{a^2}+5 \left(B(r)-1\right)\right)+36 (B(r)-1)^3=-96 \pi  \rho  r^2 (B(r)-1)^2.
\end{equation}
Applying the solution in Eq.~(\ref{eq:SolutionSpecified-B(r)}) and doing rearrangements, we arrive at the mass relation
\begin{eqnarray}\label{eq:MassRelation-1}
M = \frac{1}{36 a^2}\left\{ 2 a^2 r^2 \left(9 \kappa \mu (r\mu)^\frac{1}{3} + 8 \pi r \rho\right) + \frac{a^4 r^6 \left(15 \mu^4 - 256 \pi^2 \rho^2\right)}{\zeta(\tau)^{\frac{1}{3}}} - \zeta(\tau)^\frac{1}{3}\right\},
\end{eqnarray}
with
\begin{eqnarray}\label{eq:zeta(tau)}
\zeta(\tau) &=& -8 \pi  a^6 \rho  r^9 \left(512 \pi ^2 \rho ^2-45 \mu ^4\right)-324 a^4 \mu ^4 r^9\nonumber\\
&&+3 \sqrt{3} \sqrt{a^8 \mu ^4 r^{18} \left(125 a^4 \mu ^8+98304 \pi ^3 a^2 \rho ^3-16 \mu ^4 \left(100 \pi ^2 a^4 \rho ^2+540 \pi  a^2 \rho -243\right)\right)}.
\end{eqnarray}
Indeed the mass in Eq.~(\ref{eq:MassRelation-1}) can be recast as 
\begin{equation}\label{eq:MassRelation-2}
M = \frac{1}{3} \rho V + \frac{\kappa \mu (r \mu)^\frac{1}{3} r^2}{2} + \frac{1}{36 a^2}\Big\{\dots\Big\},
\end{equation}
where $V = \frac{4}{3} \pi r^3$ is the volume of a spherical star. The above result shows the relation between the mass and the matter/energy density of a collapsing spherical star in $R+\mu^4/R$ gravity. One notes that this relation is not naively $M=\rho V$, as it is in the case of general relativity \cite{Poisson2009}. This implies the existence of extra kinds of matter distribution inside the region $\mathcal{V}^-$ in this modified theory of gravity. {{Letting $\mu=0$ in Eqs.~(\ref{eq:MassRelation-1}) and (\ref{eq:zeta(tau)}), we get back to the desired result in general relativity (with $\sqrt[3]{-1}=-1$).}} The mass relation in Eq.~(\ref{eq:MassRelation-1}) is an important trait of the OS collapse of a spherical star in $R+\mu^4/R$ gravity when the interior and exterior regions of the star are smoothly joined across its surface. In the next section we demonstrate the kinematical properties of such surface, when it is subjected to the gravitational field of a spherical mass.

\section{Freely Falling Time-Like Surface}\label{sec:convergenceLinked}

Think of the star's surface as a time-like congruence generated by $u^\alpha$, which falls freely on the gravitational field of the star's mass $M$, in the static spacetime described by the metric (\ref{eq:SolutionGeneral}). In this section, we take care about the transverse behavior of the above congruence. This way, {{and in the context of satisfaction/violation of the SEC,}} we will be able to investigate the convergence conditions of the congruence which here, defines the OS collapse of the star's surface. We treat $\Sigma_+$ independently of its inner counterpart $\Sigma_-$. This means that we just investigate the behavior of the time-like flow $\Sigma_+$ and the evolution of its transverse subspace, in the region $\mathcal V^+$. So the geometry is supposed to be described by $g_{\alpha\beta}$. To proceed with this, let us define the tensor 
\begin{equation}\label{eq:Balphabeta-General}
B_{\alpha\beta} = u_{\alpha;\beta}
\end{equation} 
in $\mathcal{V}^+$. To talk about the transverse behavior, we note here that $B_{\alpha\beta}$ has some longitudinal components. Therefore we eliminate them by projecting the tensor onto a transverse subspace by means of a projection operator, to obtain the transverse part
\begin{equation}\label{eq:barB-General}
\bar B_{\alpha\beta} = h\indices{_\alpha^\gamma} h\indices{_\beta^\lambda} B_{\gamma\lambda},
\end{equation}
with
\begin{equation}\label{eq:halphabeta-normalized-General}
h\indices{_\alpha^\beta} = \delta_\alpha^\beta + u_\alpha u^\beta
\end{equation}
to be the projection operator. The above projector satisfies $ u^\alpha h\indices{_\alpha^\beta} = h\indices{_\alpha^\beta} u_\beta =0$ and $h\indices{^\alpha_\alpha} = 3$, therefore the transverse subspace is 3-dimensional, however it is described in a 4-dimensional spacetime structure. The transverse tensor $\bar B_{\alpha\beta}$ can itself be decomposed into symmetric and anti-symmetric parts in the following way: 
\begin{equation}\label{eq:barB-decomposed-General}
\bar B_{\alpha\beta} = \bar B_{(\alpha\beta)} + \bar B_{[\alpha\beta]} = \theta_{\alpha\beta} + \omega_{\alpha\beta},
\end{equation}
in which
\begin{equation}\label{eq:theta_alphabeta-decomposed-General}
\theta_{\alpha\beta} = \frac{1}{3} \Theta h_{\alpha\beta} + \sigma_{\alpha\beta},
\end{equation}
with $h_{\alpha\beta} = h\indices{_\alpha^\lambda} g_{\lambda\beta}$, and
\begin{subequations}\label{eq:ShearRotationExpansion-definitions}
\begin{align}
&\Theta = \bar B\indices{^\alpha_\alpha} = u\indices{^\alpha_{;\alpha}},
\label{eq:Expansion-definitions}
\\
&\sigma_{\alpha\beta} =h\indices{_\alpha^\sigma} h\indices{_\beta^\lambda} u_{(\sigma;\lambda)} - \frac{1}{3} \Theta h_{\alpha\beta},
\label{eq:Shear-definitions}
\\
&\omega_{\alpha\beta} = h\indices{_\alpha^\sigma} h\indices{_\beta^\lambda} u_{[\sigma;\lambda]},
\label{eq:Rotation-definitions}
\end{align}
\end{subequations}
are respectively the expansion scalar, and traceless shear and vorticity tensors. The above kinematical properties are involved in the evolution equation of the transverse subspace, i.e. 
\begin{equation}\label{eq:Raychaudhuri-Common}
\dot\Theta + \frac{1}{3} \Theta^2 = -\sigma_{\alpha\beta} \sigma^{\alpha\beta}
+\omega_{\alpha\beta}\omega^{\alpha\beta} - R_{\alpha\beta} u^\alpha u^\beta
+a\indices{^\alpha_{;\alpha}},
\end{equation}
known as the Raychaudhuri equation. In its original form, the Raychaudhuri equation provided a generic formulation of the gravitational attraction for a pressure-free fluid, which emphasized the repulsive gravitational energy of a positive cosmological constant \cite{Raychaudhuri1955}. In its modified form however, it is also a useful tool in inspecting the Hawking-Penrose singularity theorems \cite{Penrose1965,Hawking1965,Hawking1966} (for a good review, see Ref.~\cite{Kar2007}). In Eq.~(\ref{eq:Raychaudhuri-Common}), $a^\alpha = u\indices{^\alpha_{;\beta}} u^\beta$, and because of the peculiar choice of $h\indices{_\alpha^\beta}$ in Eq.~(\ref{eq:halphabeta-normalized-General}), the congruence is hypersurface orthogonal and therefore $\omega_{\alpha\beta} = \mathbf{0}$. The star's surface is regarded as a time-like congruence generated by the geodesic velocity 4-vector $u^\alpha$, which is characterized by Eqs.~(\ref{eq:1stJunction-11}) and (\ref{eq:2ndJunction-11}). Subsequently, we have
\begin{equation}\label{eq:u alpha-geodesic}
u^{\alpha }=\left(\frac{\tilde{E}}{A(r)}, -\sqrt{B(r) \left(\frac{\tilde{E}^2}{A(r)}-1\right)}, 0, 0\right),
\end{equation}
which satisfies $u^\alpha u_\alpha = -1 $ and $a^\alpha = \mathbf{0}$. In the spacetime described by Eqs.~(\ref{eq:SolutionGeneral}) and (\ref{eq:SolutionSpecified}), and along with the congruence generated by $u^\alpha$, then Eq.~(\ref{eq:Raychaudhuri-Common}) can be rearranged to give
\begin{equation}\label{eq:EnergySimplified}
\tilde{E}=\pm \frac{1}{2} \sqrt{\frac{-154 M^2+2 M r \left(63 \kappa  (\mu  r)^{4/3}+76\right)-3 r^2 \left(7 \kappa ^2 (\mu  r)^{8/3}+19 \kappa  (\mu  r)^{4/3}+12\right)}{r (7 M-2 r)}},
\end{equation}
{{which has the general relativistic limit $\tilde{E}_{\mathrm{GR}} = \pm\left[18-\frac{M}{2r}(13r-77M)/(2r-7M)\right]^{1/2}$.}} The above relation, retains the conformity between the junction conditions on the star's surface (giving the mass), and the satisfaction of the Raychaudhuri equation (giving the energy). In this way, one can proceed with investigating the SEC. In general relativity, the SEC concerns about the concept of an attractive gravitational force and is of great importance in focusing theorems. If this condition is satisfied, the congruence does feel an attractive gravitational field and subsequently converges~\cite{Poisson2009}. However in what follows, we exhibit some cases of violating the SEC. This means that in $R+\mu^4/R$ theory, the gravitational field may cause some repulsive issues. {{This would inevitably be interpreted as the congruence divergence. However, we argue that most of the times, the congruence is still convergent and obeys the OS collapsing regime. This would become the anomaly that we had pointed out in the introductory section; convergence within repulsive gravity in the context of an alternative theory.}}  Mathematically, the SEC declares that for the future directed vector $u^\alpha$, we have
\begin{equation}\label{eq:SEC-generalDefinition}
\left(T_{\alpha\beta} - \frac{1}{2} T g_{\alpha\beta}\right)~u^\alpha u^\beta \geq 0,
\end{equation}
where $T\equiv T\indices{^\alpha_\alpha}$. Applying this in the field equations (\ref{eq:FieldEqn}), would prove that
\begin{equation}\label{eq:SEC-mu4-1}
\left[\left(1-\frac{\mu^4}{R}\right)R_{\alpha\beta} + \frac{\mu^4}{R} g_{\alpha\beta} - \mu^4 \left(\nabla_\alpha\nabla_\beta + \frac{1}{2} g_{\alpha\beta} \Box\right) \frac{1}{R^2}\right]u^\alpha u^\beta \geq 0.
\end{equation}
{{For $\mu=0$, we get back to the famous form of the SEC, i.e. $R_{\alpha\beta} u^\alpha u^\beta \geq 0 $ in general relativity.}} Equation~(\ref{eq:SEC-mu4-1}) can be manipulated to give
\begin{eqnarray}\label{eq:SEC-mu4-2}
&&\left[\left(1-\frac{\mu ^4}{R^2}\right) G_{\alpha \beta } + 
\frac{1}{2} \left(1+\frac{\mu^4}{R^2}\right) R g_{\alpha\beta}
- \frac{3 \mu^4}{R^4} \left\{(\partial^\lambda R) (\partial_\lambda R) g_{\alpha\beta} + 2(\partial_\alpha R)(\partial_\beta R)\right\}\right.\nonumber\\
&&\left. +\frac{\mu^4}{R^3}\left(g_{\alpha\beta} \Box R + 2\partial_\alpha\partial_\beta R\right)\right] u^\alpha u^\beta \geq 0,
\end{eqnarray}
in which $G_{\alpha\beta} \equiv R_{\alpha\beta} - \frac{1}{2} R g_{\alpha\beta}$ is the Einstein tensor. Since in the static spacetime under consideration, it is $R\equiv R(r)$, the above relation is simplified to
\begin{equation}\label{eq:SEC-mu4-3}
\left(1-\frac{\mu ^4}{R^2}\right) G_{\alpha \beta } u^\alpha u^\beta
- \frac{1}{2} \left(1+\frac{\mu^4}{R^2}\right) R + \frac{3 \mu^4}{R^4} B(r) R'^2
-\frac{6\mu^4}{R^4} R'^2 (u^r)^2
-\frac{\mu^4}{R^3} B(r) R'' 
+\frac{2\mu^4}{R^3} R'' (u^r)^2 \geq 0,
\end{equation}
to obtain which, we have used $u^\alpha u_\alpha = -1$. Finally, interpolation of Eq.~(\ref{eq:u alpha-geodesic}) yields
\begin{equation}\label{eq:SEC-mu4-4}
\left(1-\frac{\mu ^4}{R^2}\right) R_{\alpha \beta } u^\alpha u^\beta
- \frac{\mu^4}{R}
 + \mu^4 B(r) \left(\frac{2\tilde E^2}{A(r)} - 3\right)\left(\frac{R''}{R^3} + \frac{2 R'^2}{R^4}\right) \geq 0.
\end{equation}
On a star's surface of fixed radius $r = r_0$, the above SEC can be specified for the spacetime given in Eqs.~(\ref{eq:SolutionGeneral}) and (\ref{eq:SolutionSpecified}). This condition is therefore characterized by $\mathcal{M}$ in Eq.~(\ref{eq:SEC-geodesic-app}) on the star's surface, in which $M_0$ and $\tilde{E}_0$ are those fixed values of the star's mass and congruence energy, which are obtained by letting  $r=r_0$ in Eqs.~(\ref{eq:MassRelation-1}) and (\ref{eq:EnergySimplified}). Note that $M_0$ contains the fixed value $a_0 = r_0/\sin\chi_0$ (because in our adopted geometric units, $\mathrm{dim}[a] = \mathrm{m}$. In all forthcoming numerical examples, we let $\chi_0=\pi/2$). We consider two examples, in order to demonstrate the satisfaction/violation of the above SEC within a range of $R+\mu^4/R$ gravity, being characterized by the parameter $\mu$. To do this, we consider a fixed sized star of a fixed density. Varying $\mu$ is equivalent to variant amounts of star's mass and congruence energy, which are given in relevant types of $R+\mu^4/R$ gravity. In this way, we can examine the SEC in those theories which are consistent with observed data, and investigate the contraction of the star's surface within them. {{In general,}} if the SEC is satisfied, we can infer that the star's surface (here known as the time-like congruence) experiences an attractive gravitational field and it subsequently converges. This can be regarded as a collapsing process, when the SEC does hold. However, if the SEC is violated, the surface is no longer confined to {{an attractive}} gravitational field of the total mass $M$; it encounters a repulsive force. Through this process, although the central mass is contracting (affected by the FLRW spacetime inside $M$), the outer surface {{is indeed supposed to hold up or diverge. However as we will see in what follows, this is not always the case in $R+\mu^4/R$ gravity; we can expect collapse in repulsive situations.}}\\

\begin{figure}[t]
\center{\includegraphics[width=7cm]{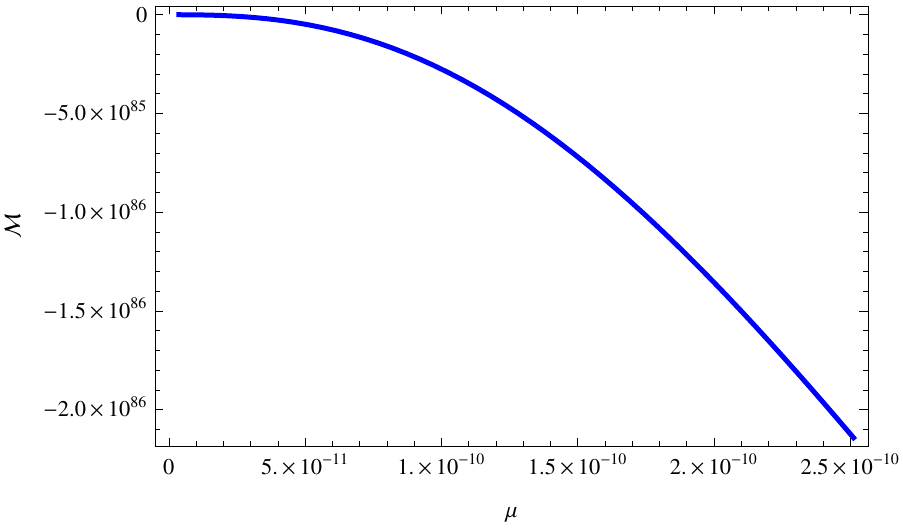}~(a)
\hfil
\includegraphics[width=7cm]{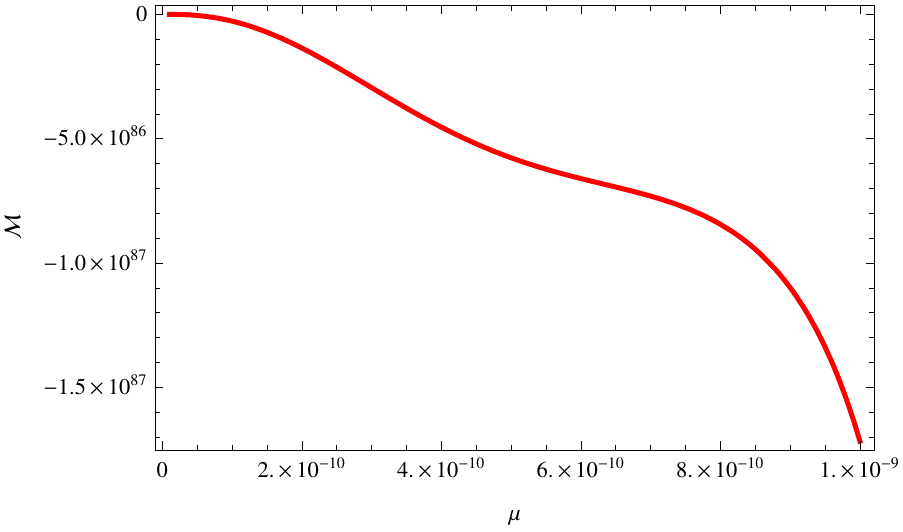}~(b)
\caption{\label{fig:CongruenceSun-geodesic} The behavior of $\mathcal{M}$ for a congruence moving on the Sun, in terms of $\mu$. The ranges of $\mu$ have been designated as (a) $10^{-13} \leq \mu \leq 10^{-9.6}$  and (b) $10^{-11} \leq \mu \leq 10^{-9}$. As it is seen, for the whole range and on the star's surface $r_0$, it is $\mathcal{M} <0$ and the SEC is violated. }}
\end{figure}

For the first example, we consider the Sun as the generator of the exterior geometry, in $R+\mu^4/R$ gravity. In accordance with the necessary values in Eq.~(\ref{eq:SEC-geodesic-app}), this star is characterized by $r_0 = r_\odot = 6957 \times 10^5 ~\mathrm{m}$ and $\rho = \rho_\odot = 1410 ~\frac{\mathrm{kg}}{\mathrm{m^3}} = 1.04705 \times 10^{-24} ~ \mathrm{m^{-2}}$ \cite{Zombeck1990}. Based on these values and according to Eq.~(\ref{eq:MassRelation-1}), only $\mu\sim10^{-11}~ \mathrm{m^{-1}}$ can give the Sun's mass which is approximately $M_0 = M_\odot \approx 2\times 10^{30}~ \mathrm{kg} = 1485.18 ~\mathrm{m}$. However in Fig.~\ref{fig:CongruenceSun-geodesic}, we have plotted $\mathcal{M}$ for a wider range for $\mu$. As it is seen, no specified $R+\mu^4/R$ theory will obey the SEC. This means that for a typical star like the Sun, the theory predicts a {{violation of the SEC for}} the star's surface; this {{however is done within a collapsing regime, because using Eq.~(\ref{eq:Expansion-definitions}) we have
\begin{eqnarray}\label{eq:Expansion-extended}
\Theta &=& -\frac{14 \left(-2 M_0+\kappa  {r_0} (\mu  {r_0})^{4/3}+{r_0}\right)^2}{{r_0}^2 (2 {r_0}-7 M_0)}\nonumber\\
&&-\left[-196 {M_0}^3+2 {M_0}^2 {r_0} \left(154 \kappa  (\mu  {r_0})^{4/3}+157\right)-M_0 {r_0}^2 \left(147 \kappa ^2 (\mu  {r_0})^{8/3}+315 \kappa  (\mu  {r_0})^{4/3}+164\right)\right.\nonumber\\
&&\left.+{r_0}^3 \left(21 \kappa ^3 \mu ^4 {r_0}^4+72 \kappa ^2 (\mu  {r_0})^{8/3}+79 \kappa  (\mu  {r_0})^{4/3}+28\right)\right]^{-1/2}\left[{r_0} (2 {r_0}-7 M_0) \left\{{r_0} (3 \kappa  (\mu  {r_0})^{4/3}+4)\right.\right.\nonumber\\
&&\left.\left.-8 M_0\right\}\right]^{1/2},
\end{eqnarray}
on the star's shell. This is always a negative valued scalar for those $\mu$ in Fig.~\ref{fig:CongruenceSun-geodesic}. This therefore exhibits an anomalous behavior of the gravitational system. Note that, in the general relativistic limit ($\mu=0$), we have $R_{\alpha\beta} u^\alpha u^\beta = 0$; satisfaction of the SEC, and 
\begin{eqnarray}\label{eq:Expansion-GR}
\Theta_{\mathrm{GR}} = - \frac{14 ({r_0}-2 M_0)^2}{{r_0}^2 (2 {r_0}-7 M_0)} \left(\frac{49 {M_0}^2-54 M_0 {r_0}+14 {r_0}^2}{r_0(4 r_0 -14 {M_0})}\right)^{-1/2},
\end{eqnarray}
which because of its negativity, is in agreement with the classical collapsing criteria.
}}
\\

\begin{figure}[top]
\center{\includegraphics[width=7cm]{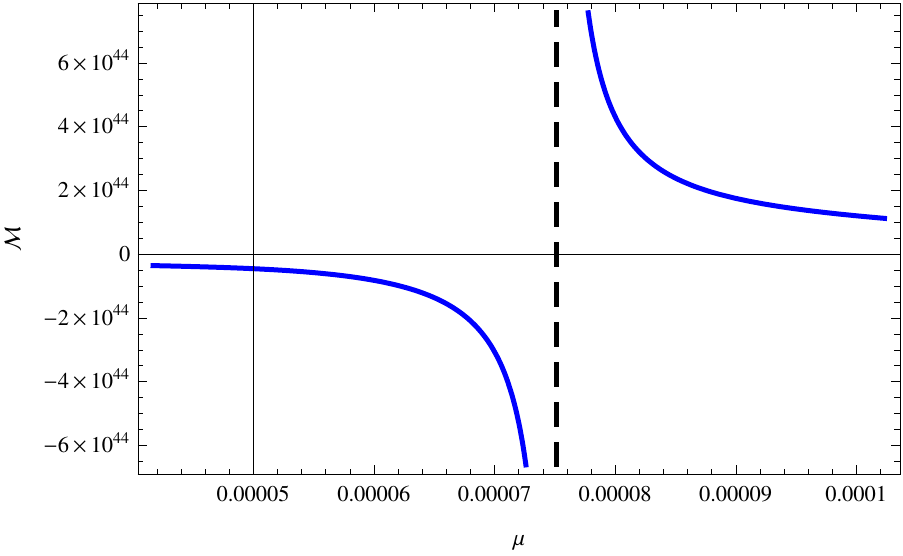}
\caption{\label{fig:CongruenceNeutron-geodesic} The behavior of $\mathcal{M}$ for a congruence moving on a typical neutron star. The range of $\mu$, varies the mass of the star within $1.1 M_\odot \lessapprox M_0 \lessapprox 3 M_\odot $. It is seen that for the lower limit, the SEC is violated whereas for the upper limit, it is satisfied. This implies a congruence convergence for more massive stars {{within attractive gravity}}. }}
\end{figure}

For the second example, we turn to a much more massive case, i.e. a typical neutron star. Indeed neutron stars have actually passed a gravitational collapsing era and have reached a state in which the fermionic pressure is able to stabilize the star's state, by overcoming the gravitational attraction. We examine the stability of a neutron star in $R+\mu^4/R$ gravity by observing the behavior of its surface, through inspection of the SEC. The size of a common neutron star is about $r_0 \approx 11 ~\mathrm{km}$  and $\rho = 4.8 \times 10^{17}~\frac{\mathrm{kg}}{\mathrm{m^3}}$ and a typical mass range, between $1.1M_\odot$ and $3 M_\odot$ \cite{Glendenning2000}. Therefore the mass relation (\ref{eq:MassRelation-1}) implies the domain $10^{-4.38} \leq \mu \leq 10^{-3.99}$ to cover the above mass range. The behavior of $\mathcal{M}$ has been plotted in Fig.~\ref{fig:CongruenceNeutron-geodesic}.
The illustration shows that lighter stars (about $1.1 M_\odot$) are still incapable of attracting their surface during the contraction of their interior segment. Therefore, as in the Sun's case, for these stars the surface do not experience an attractive gravitational field. However for the heavier ones (those about $3 M_\odot$), the SEC is satisfied, the surface is trapped in an attractive gravitational field. {{
The expansion in Eq.~(\ref{eq:Expansion-extended}) appears to be still of negative values for those $\mu$ in Fig.~\ref{fig:CongruenceNeutron-geodesic}. Therefore one can infer that the convergence is done within repulsive gravity of the exterior region $\mathcal{V}^+$, in the case of smaller neutron stars. This anomalous behavior is indeed different from the collapsing criteria defined in the context of general relativity. For heavier stars, such a collapse is done within attractive gravity.}} \\

The above results, indicate that even for smoothly joined regions, different versions of the theory will imply different {{behaviors}} for the gravitational systems. However one significant consequence is that the {{normal}} gravitational collapse will occur {{only}} in extremely heavy stars. {{This consequence is consistent with general relativity's predictions.}} In the next section, we deal with a totally different realm, i.e. disconnected interior and exterior regions. This exhibits {{normal}} congruence convergence for light stars in $R+\mu^4/R$ gravity.

\section{Violation of the First Junction Condition and the Consequent Anomalies} \label{sec:convergenceNon-normalized}

So far, we have taken care about those spherically symmetric gravitational systems, where the space-like hypersurface served as a joining boundary for the interior and exterior regions. However this question can be raised as to what would happen if such boundary were a separation of two regions, showing considerable differences between their geometrical properties. In the OS collapse, as discussed in Sec.~\ref{sec:massOS}, the whole process is founded on the smooth geometrical connection which is established by using the first junction condition $[\gamma_{ab}] = \mathbf{0}$. As it has been pointed out, the main aim of the first junction condition is to create a union of exact solutions to the gravitational field equations. In this section, we assume that this condition is violated at some levels. This means that the induced metrics $\gamma^\pm_{ab}$ are not completely joined along $\Sigma_-\longleftrightarrow\Sigma_+$, according to the surface structure in Fig.~\ref{fig:Sigma-DoubleLayer}, and lose their connection. In $R+\mu^4/R$ gravity, such situation may be encountered for example when the interior region scale factor does not satisfy the Friedmann equation (\ref{eq:Friedmann-1}). Therefore, the union of spacetimes in the regions $\mathcal{V}^\pm$ cannot generate a solid solution to the $R+\mu^4/R$ field equations and the physical nature of the interior and exterior regions of the star exhibits significant differences. We display this violation by the following condition:
\begin{equation}\label{eq:1stJunction-Violation-general}
\left[\gamma_{ab}\right] = \mathrm{diag} \left(\Phi(r), 0, 0\right),
\end{equation}
in which $\Phi(r)$ is allowed to have roots and therefore, can be either positive or negative. The induced metrics $\gamma^\pm_{ab}$ in Eq.~(\ref{eq:IntrinsicMetrics - +-}), now give
\begin{equation}\label{eq:1stJunction-Violation-11,22,33}
\left[\gamma_{11}\right] ={} \Phi(r) ~~\Longrightarrow~ A(r) \dot t = \sqrt{\frac{A(r)}{B(r)} \left(B(r) \Phi(r) + B(r) + \dot r^2\right)}.
\end{equation}
Accordingly, the tangential congruence generator of the exterior layer $\Sigma_+$ is derived as
\begin{equation}\label{eq:u alpha-nonGeodesic}
u^\alpha = \left(\dot t, -\sqrt{A(r) B(r) \dot t^2 - B(r)\left(\Phi(r) + 1\right)}, 0, 0\right).
\end{equation}
A specific choice can be $\dot t = 1$ which is adopted here. As we are in the region $\mathcal{V}^+$, this means $u^\alpha u_\alpha = -\left(\Phi+1\right)$ and $a^\alpha =\left(-(A'/A) \sqrt{B (A-\Phi-1)}, B(A'-1/2 ~\Phi'), 0, 0\right) $. We {{seek}} for appropriate conditions on $\Phi(r)$, to make it possible to look into the SEC and its relevance to congruence convergence and subsequent gravitational collapse of the outer layer of the star's surface. As in the previous section, we once again exploit the evolution equation of the transverse subspace (or the Raychaudhuri equation) to proceed with this task. However, since the tangential vector $u^\alpha$ is not normalized to $-1$ (or any other constant values), the projection operator in Eq.~(\ref{eq:halphabeta-normalized-General}), now takes the form
\begin{equation}\label{eq:halphabeta-NONnormalized-General}
h\indices{_\beta^\alpha} = \delta_\beta^\alpha - \frac{u_\beta u^\alpha}{u_\gamma u^\gamma},
\end{equation}
which still satisfies $u^\beta h\indices{_\beta^\alpha} = h\indices{_\beta^\alpha} u_\alpha = 0$ and $h\indices{^\alpha_\alpha} =3$. The evolution of the resultant transverse subspace, obtained by doing projections, is therefore given by the generalized Raychaudhuri equation for non-normalized tangential vectors (see Appendix \ref{app:DerivationRaychaudhuri})
\begin{eqnarray}\label{eq:Raychaudhuri-NonNormalized}
\dot \Theta + \frac{1}{3} \Theta^2 & = & -\sigma_{\alpha\beta} \sigma^{\alpha\beta} + \omega_{\alpha\beta} \omega^{\alpha\beta} + h\indices{_\beta^\alpha} a\indices{^\beta_{;\alpha}}
-h\indices{_\beta^\alpha} R\indices{^\beta_{\mu\alpha\nu}} u^\mu u^\nu\nonumber\\
&& + \frac{u^\alpha}{2(u_\lambda u^\lambda)^2}\left[2 a_\gamma u^\gamma \left(2 a_\alpha + (u^\mu u_\mu)_{;\alpha}\right) + 
\left(a_\alpha u^\mu + u_\alpha a^\mu\right) \left(2 a_\mu + (u^\nu u_\nu)_{;\mu}\right)
\right]\nonumber\\
&&-\frac{1}{u_\lambda u^\lambda} \left[2 a^\mu a_\mu + (u^\mu u_\mu)_{;\alpha} a^\alpha \right] - \left(\frac{2  a_\gamma u^\gamma}{u_\lambda u^\lambda}\right)^2,
\end{eqnarray}
with 
\begin{subequations}\label{eq:Raychaudhuri-expansion,shear,vorticity-NonNormalized}
\begin{align}
&\Theta = u\indices{^\alpha_{;\alpha}} - \frac{u^\alpha a_\alpha}{ u_\lambda u^\lambda},
\label{eq:Raychaudhuri-expansion-NonNormalized}
\\
&\sigma_{\alpha\beta} = h\indices{_\alpha^\gamma} h\indices{_\beta^\lambda} u_{(\gamma;\lambda)} - \frac{1}{3} h_{\alpha\beta} \Theta,
\label{eq:Raychaudhuri-shear-NonNormalized}
\\
&\omega_{\alpha\beta} = h\indices{_\alpha^\gamma} h\indices{_\beta^\lambda} u_{[\gamma;\lambda]},
\label{eq:Raychaudhuri-vorticity-NonNormalized}
\end{align}
\end{subequations}
according to Eq.~(\ref{eq:halphabeta-NONnormalized-General}). For any normalized vectors, satisfying $u^\alpha u_\alpha = -1$, the projection operator (\ref{eq:halphabeta-NONnormalized-General}) reduces to that in Eq.~(\ref{eq:halphabeta-normalized-General}) and we get back to the usual form of the Raychaudhuri equation in Eq.~(\ref{eq:Raychaudhuri-Common}) (it is also true for any $u^\alpha u_\alpha = \mathrm{const.}$ See Appendix \ref{app:DerivationRaychaudhuri}). Note that, the above generalized Raychaudhuri equation for non-normalized tangential vectors, is different from that which has been introduced in Ref.~\cite{Abreu2011} and has different applications. First of all, the transverse subspace of which the kinematical properties are investigated in this paper, was obtained by means of the transverse projector (\ref{eq:halphabeta-NONnormalized-General}) and is therefore different from that which has been used in Ref.~\cite{Abreu2011}. Secondly, our treatment method is completely 4-dimensional and the transverse subspace does not coincide with $\Sigma$ (it is indeed {{transverse}} to it). Hence, we do not encounter the necessity of inclusion of the extrinsic curvature of $\Sigma$. \\

In the $\mathcal{V}^+$ and for a congruence generated by $\mathbf{u}$ in Eq.~(\ref{eq:u alpha-nonGeodesic}), the above generalized Raychaudhuri equation gives the following differential equation:
\begin{equation}\label{eq:DifferentialEqn-Phi(r)}
2 \left(\Phi (r)+1\right) A'(r)-A(r) \Phi '(r) = 0,
\end{equation}
which has the solution
\begin{equation}\label{eq:DifferentialEqn-Phi(r)-Solution}
\Phi(r) = -1 + A(r)^2.
\end{equation}
The above function has one positive-valued degenerate root at $\mathcal{R} = 2 \left(\frac{4 M^3}{27 \kappa^3 \mu^4}\right)^{1/7}$, implying that $\Phi(r)$ can be negative or positive. However the expression in Eq.~(\ref{eq:DifferentialEqn-Phi(r)-Solution}) yields $u^\alpha u_\alpha = -A(r)^2$ which guarantees that the congruence remains time-like. Note that, at points on which $A(r) = 1$, the vanishing of $\Phi(r)$ results in $u^\alpha = (1, 0, 0, 0)$ which is the velocity of a comoving observer with $\Sigma_+$. This, as it is expected, implies the smooth connection retrieval. Once again we turn to the SEC and its impacts on possible ranges for $\mu$, which do correspond to gravitational collapse in certain classes of $R+\mu^4/R$ theory. The SEC relation in Eq.~(\ref{eq:SEC-mu4-2}), for $R = R(r)$ and specific to the tangential vector (\ref{eq:u alpha-nonGeodesic}) and together with Eq.~(\ref{eq:DifferentialEqn-Phi(r)-Solution}), takes the form 
\begin{equation}\label{eq:SEC-NonNormalized}
\left(1-\frac{\mu ^4}{R^2}\right) R_{\alpha\beta} u^\alpha u^\beta
-\frac{\mu^4}{R} A(r)^2
+ \frac{\mu ^4}{R^3} A(r) B(r) \left[3 A(r) -2\right] \left\{\frac{3 R'^2}{R} - R''\right\} \geq 0.
\end{equation}
Analogously to the previous section, now in order to investigate the possibility of collapsing of a star's surface which has lost the connection with its interior region, we apply the above SEC on a star of the fixed radius $r_s$ and fixed known mass $M_s$. The values of Eq.~(\ref{eq:SolutionSpecified}), if applied to the above condition, will result in the algebraic expression of $\mathcal{M}$ in Eq.~(\ref{eq:SEC-nongeodesic-app}).
 
As before, we intend to find out which ranges amongst $R+\mu^4/R$ classes can serve as representatives of exterior geometries of \textit{collapsing} stars. Note that since the junction condition is violated, there is no need to obtain the mass of the star in connection with its exterior geometry. We therefore consider a spherical star with definite mass $M_s$ and radius $r_s$ and demonstrate the behavior of the gravitational field, acting on a time-like flow while evolving in the exterior geometry of such a mass. This would be the realm of applying the Raychaudhuri equation for a time-like congruence in $\mathcal{V}^+$. We exemplify this by considering the Sun and a typical neutron star as the sources of gravity.\\

\begin{figure}[t]
\center{\includegraphics[width=7cm]{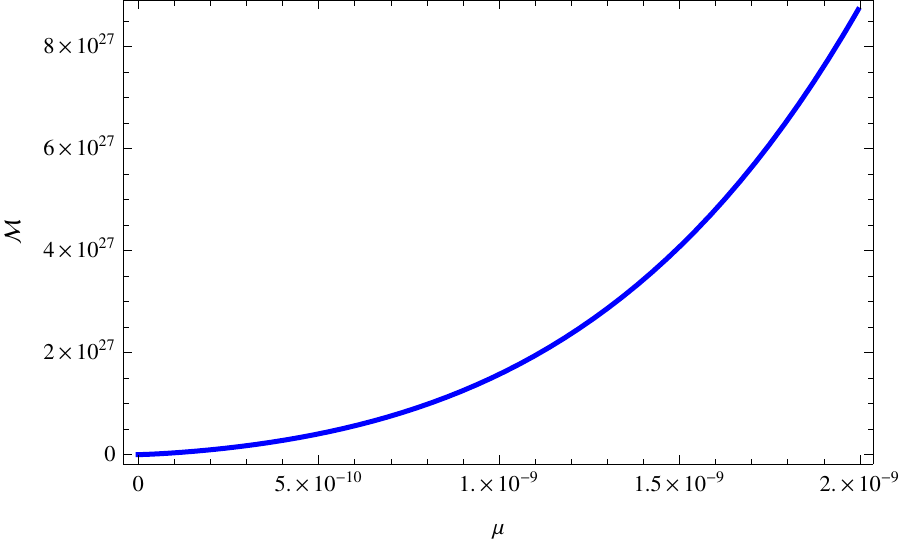}
\caption{\label{fig:CongruenceSun-Nongeodesic} The behavior of $\mathcal{M}$ for a congruence moving on the Sun, in terms of $\mu$. The range of $\mu$ has been designated as $0 ~{{<}}~ \mu \leq 10^{-8.7}$. As it is seen, for the whole range and on the star's surface $r_s$, it is $\mathcal{M} >0$ and the SEC is satisfied. }}
\end{figure}

The Sun, as the first example, is characterized by $r_s = 6957\times 10^5 ~\mathrm{m}$ and $M_s = 1485.18 ~\mathrm{m}$. In Fig.~\ref{fig:CongruenceSun-Nongeodesic}, we have plotted the behavior of $\mathcal{M}$ with respect to $\mu$, for a congruence moving on $r = r_s$.  We can observe that huge positive values of $\mathcal{M}$ are given for rather small values of $\mu$. This means that in all classes of $R+\mu^4/R$ theory of gravity and when the first junction condition is violated, the outer shell of an object even like the Sun, will experience {{
an attractive gravitational field. As we recall, this was not the case in the smoothly joined regions, in the previous section. However, when the congruence expansion is considered, the actual anomaly rises. On the massive object, Eq.~(\ref{eq:Raychaudhuri-expansion-NonNormalized}) together with Eq.~(\ref{eq:u alpha-nonGeodesic}), give 
\begin{eqnarray}\label{eq:Expansion-Violated-First}
\Theta &=& {r_s}^{-5/2}{\left(-2 M_s+\kappa  {r_s} (\mu  {r_s})^{4/3}+{r_s}\right) \left(2 \kappa {r_s} (\mu  {r_s})^{4/3}-3 M_s\right) \left[{r_s} \left(3 \kappa  (\mu  {r_s})^{4/3}+4\right)-8 M_s\right]}{}\nonumber\\
&&\times\left[128 {M_s}^3-32 {M_s}^2 {r_s} \left(5 \kappa  (\mu  {r_s})^{4/3}+4\right)+2 M_s {r_s}^2 \left(33 \kappa ^2 (\mu  {r_s})^{8/3}+52 \kappa  (\mu  {r_s})^{4/3}+16\right)\right.\nonumber\\
&&\left.-3 \kappa  \mu  {r_s}^4 \left(3 \kappa ^2 \mu ^3 {r_s}^3+7 \kappa  (\mu  {r_s})^{5/3}+4 \left(\mu  {r_s}\right)^{1/3}\right)\right]^{-1/2},
\end{eqnarray}
which is indeed of positive values for those $\mu$ in Fig.~\ref{fig:CongruenceSun-Nongeodesic}. This shows that in $R+\mu^4/R$ gravity, the stars below the Chandrasekhar mass ($\sim 1.44~M_\odot$), who have lost the connection between their interior and exterior geometries, do not obviate the necessities of a vigorous OS collapse. This exhibits an extreme difference with the same stars who satisfy the first junction condition. Note that, the general relativistic limit ($\mu=0$) implies
\begin{eqnarray}\label{eq:Expansion-Violated-First-GR}
\Theta_{\mathrm{GR}} = - \frac{3 ({r_s}-2 M_s)}{\sqrt{2}~ r_s} \left[\frac{M_s }{{r_s}}\right]^{1/2},
\end{eqnarray}
which is always negative for the mass and radius of Fig.~\ref{fig:CongruenceSun-Nongeodesic} (also the general relativistic SEC still gives $R_{\alpha\beta} u^\alpha u^\beta = 0$), and we therefore have a normal OS collapse in an attractive gravitational field, for all cases. The above results show the crucial impacts of the first junction condition; if it is violated for small stars, we can expect a congruence divergence (repulsion of the outer shell) within attractive gravity.
}}
Whether or not they make any sense, these results of course belong to the $R+\mu^4/R$ theory. However unless thorough observations have been done along with the satisfaction/violation of the first junction condition in various types of stars, we cannot be sure that such {{phenomena are}} completely impossible in smaller stars, because its availability relies crucially on the first junction condition in specific exterior geometries and also the way we look into the problem. Note that, for the range of $\mu$ which has been considered in the depiction of Fig.~\ref{fig:CongruenceSun-Nongeodesic}, we always have $r_s \ll \mathcal{R}$. Therefore we can be assured that none of the $R+\mu^4/R$ classes are capable of retrieving the first junction condition on the Sun's surface.\\

For a neutron star of mass $M_s = 4455.54 ~\mathrm{m}$ and radius $11\times10^3~\mathrm{m}$, we can observe two class ranges for the $R+\mu^4/R$ gravity, each of which implying a totally different fate for the star (see Fig.~\ref{fig:CongruenceNeutron-Nongeodesic}):
\begin{figure}[t]
\center{\includegraphics[width=7cm]{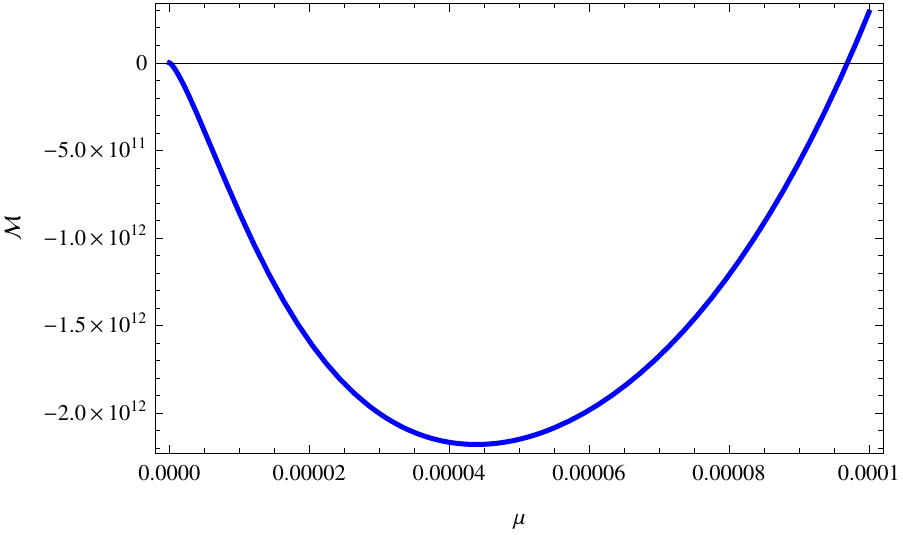}~(a)
\hfil
\includegraphics[width=7cm]{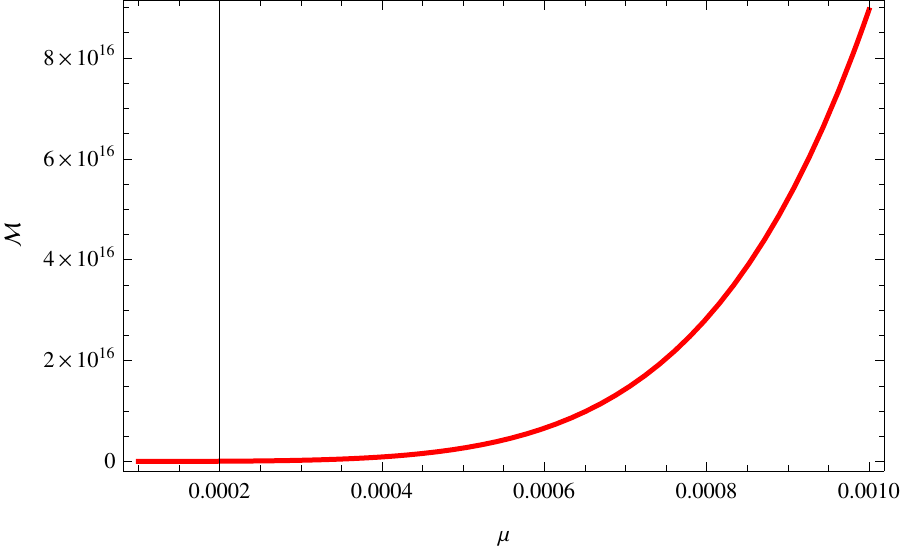}~(b)
\caption{\label{fig:CongruenceNeutron-Nongeodesic} The behavior of $\mathcal{M}$ for a congruence moving on a typical neutron star of mass $M_s = 3 M_{\odot}$, in terms of $\mu$. The ranges of $\mu$ have been designated as (a) $0 ~{{<}}~ \mu \leq 10^{-4}$  and (b) $10^{-4} \leq \mu \leq 10^{-3}$. These two cases anticipate completely different effects of gravity on the congruence. }}
\end{figure}
\begin{itemize}

\item{For $0 ~{{<}}~ \mu \leq 10^{-4}$, it is  $\mathcal{M}<0$, {{$\Theta<0$}} and therefore, gravitational collapse {{within repulsive gravity,}} because of violation of the SEC.  }

\item{For $\mu > 10^{-4}$, it is $\mathcal{M} \geq 0$, {{$\Theta > 0$}}  and the inevitable {{shell repulsion within attractive gravity}}. This category itself is divided into two parts:  $10^{-4} < \mu  \leq 2\times 10^{-4}$ corresponds to regions in which $r_s<\mathcal{R}$, whereas for
$\mu > 2 \times10^{-4}$, the radial range includes the value $\mathcal R$. {{In this range, we encounter a divergence for $\Theta$ in Eq.~(\ref{eq:Expansion-Violated-First}) which indicates a congruence convergence.}} Therefore at some distances, the gravitational collapse can be a result of the junction condition retrieval. }

\end{itemize}
The above results indicate that for larger massive sources, the talent of creating extremely attractive gravitational fields, still depends on the choice of $\mu$ and subsequently on the chosen type of $R+\mu^4/R$ theory. It has also been pointed out that unlike the previous example, the violation of the first junction condition in such cases, will not necessarily lead to gravitational convergence of the star's surface and one has to be specific in choosing the approach to the problem.

\section{Conclusion}\label{sec:conclusion}

The advantageous influence of smooth joining across a time-like hypersurface is not condensed in formulating the energy contents of the gravitational source and the relations to its enveloping spacetime; it also provides the chance of reproducing consistent results with well-known observations. This therefore becomes of great help in testing extended theories of gravity and their astrophysical predictions. The elaboration of such examinations can be done through studying the OS collapse in these theories. In this paper as well, we discussed the collapsing of a spherical star and the convergence criteria of its time-like surface in an extended theory, namely the $R+\mu^4/R$ gravity. We first pursued the standard procedure in which, the smooth joining across the star's surface holds by relying on the first and second junction conditions. In the same way as in the original approach to the OS collapse, we constructed a framework to find a relation between the mass content of a collapsing star and its density. The results argue that this relation does not coincide with that in general relativity, as it possesses extra terms and consequently, extra forms of mass/energy distribution. However, whether or not such extra terms appear, it should be noted that the mass relations reflect the formation of a union of exact solutions to the considered gravitational theory. For the $R+\mu^4/R$ gravity, the enveloping spacetime (as the star's exterior geometry) was supposed to be the spherically symmetric solution to the weak field limit of the theory, which shows small perturbations on the Schwarzschild spacetime. Such perturbations are hence responsible for deviations from general relativity's predicted mass distribution. However, the importance of these perturbations becomes more apparent once we turn to the evolution of the time-like boundary of the star. Maintaining the smooth joining across this boundary, we found that the parameter $\mu$ of the theory, performs a crucial role in anticipating the surface's evolution. The relevant analysis was highly based on studying the SEC on different types of stars and restricting $\mu$ into some specific ranges. We showed that, putting aside the concrete predictions of the theory for a star like the Sun, various ranges of $\mu$ will result in theories which give totally apposite {{collapsing conditions}} for heavier stars. Accordingly, we could create two different categories of the theory, each of which corresponds to {{satisfaction/violation of the SEC}}. This therefore emphasizes the importance of the adoption of the approach (i.e. a specified range in the theory) in investigating the possibility of the gravitational collapse, even when the smooth joining does hold. The situation becomes more convoluted when the star loses the connection with its exterior geometry. As a first result, the union of interior and exterior geometries do not constitute an exact solution to the gravitational field equations. This manifests itself in violation of the first junction condition. In the context of $R+\mu^4/R$ theory, we demonstrated the cruciality of the choice of $\mu$, by studying the behavior of the SEC and the characteristics of the time-like boundary. The impacts of the violation of the first junction condition, was proved to resulting in non-normalized congruence generators. Applying a generalized form of the Raychaudhuri equation, we exemplified a specific kind of violation. This example once again was used for the Sun and a typical neutron star. The illustrations showed that if the smooth joining is lost at some levels, even smaller stars may experience {{an attractive gravitational field, but a shell repulsion}}. This unexpected phenomenon appears to be available in the whole ranges of $\mu$. However for heavier stars, we still encountered two categories, giving a division within the theory which separated two completely opposite predictions. In the quest for finding reliable alternative gravitational theories, we therefore conclude that it is of benefit to look for systems which separate two geometrically disconnected regions. As it was discussed in this paper, if such systems do exist, we then may observe totally unexpected phenomena, such as {{repulsion}} in systems of smaller mass contents. We have also highlighted the importance of the choice of approach. For the theory we have considered in this paper, this importance goes to the specification of the parameter $\mu$. 

\textit{We would like to thank the referee, whose valuable comments illuminated the way to enriching the discussion.}

\appendix

\section{The Algebraic Expression of the SEC for the Geodesic Congruence}

\begin{eqnarray}\label{eq:SEC-geodesic-app}
\mathcal{M} & \doteq & 144 \mu ^2 {r_0}^2 \left[-2 M_0+\kappa  {r_0} (\mu  {r_0})^{4/3}+{r_0}\right] \left[{r_0} \left(3 \kappa  (\mu  {r_0})^{4/3}+4\right)-8 M_0\right]^2\nonumber\\
&& \times\left[5152 {M_0}^3 ({\mu  {r_0}})^{1/3}-8 {M_0}^2 {r_0} ({\mu  {r_0}})^{1/3} \left(567 \kappa  (\mu  {r_0})^{4/3}+956\right)\right.\nonumber\\
&& -3 {r_0}^3 \left(105 \kappa ^3 \mu ^4 {r_0}^4 ({\mu  {r_0}})^{1/3}+390 \kappa ^2 \mu ^3 {r_0}^3+480 \kappa  (\mu  {r_0})^{5/3}+224 ({\mu  {r_0}})^{1/3}\right)\nonumber\\
&& \left.+M_0 {r_0}^2 \left(2331 \kappa ^2 \mu ^3 {r_0}^3+5400 \kappa  (\mu  {r_0})^{5/3}+3952 ({\mu  {r_0}})^{1/3}\right)\right]^2\nonumber\\
&&+288 \mu ^2 {r_0}^2 \left(-2 M_0+\kappa  {r_0} (\mu  {r_0})^{4/3}+{r_0}\right) \left({r_0} \left(3 \kappa  (\mu  {r_0})^{4/3}+4\right)-8 M_0\right)\nonumber\\
&&\times\left({r_0} \left(-4 {\tilde{E}_0}^{~2}+3 \kappa  (\mu  {r_0})^{4/3}+4\right)-8 M_0\right) \left[5152 {M_0}^3 ({\mu  {r_0}})^{1/3}-8 {M_0}^2 {r_0} ({\mu  {r_0}})^{1/3} \left(567 \kappa  (\mu  {r_0})^{4/3}+956\right)\right.\nonumber\\
&&- 3 {r_0}^3 \left(105 \kappa ^3 \mu ^4 {r_0}^4 ({\mu  {r_0}})^{1/3}+390 \kappa ^2 \mu ^3 {r_0}^3+480 \kappa  (\mu  {r_0})^{5/3}+224 ({\mu  {r_0}})^{1/3}\right)\nonumber\\
&& \left.+ M_0 {r_0}^2 \left(2331 \kappa ^2 \mu ^3 {r_0}^3+5400 \kappa  (\mu  {r_0})^{5/3}+3952 ({\mu  {r_0}})^{1/3}\right)\right]^2 
-24 (\mu  {r_0})^{7/3} \left(-2 M+\kappa  {r_0} (\mu  {r_0})^{4/3}+{r_0}\right)\nonumber\\
&& \times\left[644 {M_0}^2 ({\mu  {r_0}})^{1/3}-546 \kappa  M_0 {r_0} (\mu  {r_0})^{5/3}-664 M_0 {r_0} ({\mu  {r_0}})^{1/3}+105 \kappa ^2 \mu ^3 {r_0}^5+270 \kappa  {r_0}^2 (\mu  {r_0})^{5/3}+168 {r_0}^2 ({\mu  {r_0}})^{1/3}\right]\nonumber\\
&&\times\left({r_0} \left(3 \kappa  (\mu  {r_0})^{4/3}+4\right)-8 M\right)^2
\left[206080 {M_0}^4-960 {M_0}^3 {r_0} \left(343 \kappa  (\mu  {r_0})^{4/3}+428\right)+96 {M_0}^2 {r_0}^2 \left(945 \kappa ^2 (\mu  {r_0})^{8/3}\right.\right.\nonumber\\
&&\left. +4570 \kappa  (\mu  {r_0})^{4/3}+3176\right) + 3 {r_0}^4 \left(1575 \kappa ^4 \mu ^5 {r_0}^5({\mu  {r_0}})^{1/3}+7590 \kappa ^3 \mu ^4 {r_0}^4+14040 \kappa ^2 (\mu  {r_0})^{8/3}+13344 \kappa  (\mu  {r_0})^{4/3}+4480\right) \nonumber\\
&& - \left.4 M_0  {r_0}^3 \left(10899 \kappa ^3 \mu ^4 {r_0}^4+36720 \kappa ^2 (\mu  {r_0})^{8/3}+57840 \kappa  (\mu {r_0})^{4/3}+26240\right)\right]
-48 (\mu  {r_0})^{7/3} \left(-2 M_0+\kappa  {r_0} (\mu  {r_0})^{4/3}+{r_0}\right)\nonumber\\
&&\times\left[ 644 {M_0}^2 ({\mu  {r_0}})^{1/3}-546 \kappa  M_0 {r_0} (\mu  {r_0})^{5/3}-664 M_0 {r_0} ({\mu  {r_0}})^{1/3}+105 \kappa ^2 \mu ^3 {r_0}^5+270 \kappa  {r_0}^2 (\mu  {r_0})^{5/3}+168 {r_0}^2 ({\mu  {r_0}})^{1/3}\right]\nonumber\\
&&\times \left({r_0} \left(3 \kappa  (\mu  {r_0})^{4/3}+4\right)-8 M_0\right)
\left({r_0} \left(-4 {\tilde{E}_0}^{~2}+3 \kappa  (\mu  {r_0})^{4/3}+4\right)-8 M_0\right)
\left[206080 {M_0}^4-960 {M_0}^3 {r_0} \left(343 \kappa  (\mu  {r_0})^{4/3}+428\right)\right.\nonumber\\
&& + 96 {M_0}^2 {r_0}^2 \left(945 \kappa ^2 (\mu  {r_0})^{8/3}+4570 \kappa  (\mu  {r_0})^{4/3}+3176\right)\nonumber\\
&&+3 {r_0}^4 \left(1575 \kappa ^4 \mu ^5 {r_0}^5 ({\mu  {r_0}})^{1/3}+7590 \kappa ^3 \mu ^4 {r_0}^4+14040 \kappa ^2 (\mu  {r_0})^{8/3}+13344 \kappa  (\mu  {r_0})^{4/3}+4480\right)\nonumber\\
&& - \left. 4 M_0 {r_0}^3 \left(10899 \kappa ^3 \mu ^4 {r_0}^4+36720 \kappa ^2 (\mu  {r_0})^{8/3}+57840 \kappa  (\mu  {r_0})^{4/3}+26240\right)\right]\nonumber\\
&& + \left(- 8 M_0 + {r_0} \left(3 \kappa  (\mu  {r_0})^{4/3}+4\right)\right)^{-2}  4 \kappa  \mu  {r_0}^2 
\left[644 {M_0}^2 ({\mu  {r_0}})^{1/3}-664 M_0 {r_0} ({\mu  {r_0}})^{1/3}+105 \kappa ^2 \mu ^3 {r_0}^5 +168 {r_0}^2 ({\mu  {r_0}})^{1/3}\right.\nonumber\\
&& +\left. 270 \kappa  {r_0}^2 (\mu  {r_0})^{5/3}-546 \kappa  M_0 {r_0} (\mu  {r_0})^{5/3}\right]^2 \left[ -8 \left(7 {\tilde{E}_0}^{~2}+45\right) M_0 {r_0} ({\mu  {r_0}})^{1/3}+336 {M_0}^2 ({\mu  {r_0}})^{1/3}+63 \kappa ^2 \mu ^3 {r_0}^5\right.\nonumber\\
&&\left. + 2 {r_0}^2 ({\mu  {r_0}})^{1/3} \left(8 {\tilde{E}_0}^{~2}-147 \kappa  \mu  M_0 ({\mu  {r_0}})^{1/3}+48\right)+156 \kappa  {r_0}^2 (\mu  {r_0})^{5/3}\right] \left[8 {M_0}^2 \left(24 \mu {r_0}-161 \kappa  ({\mu  {r_0}})^{1/3}\right)\right.\nonumber\\
&& -4 M_0 {r_0} \left(-273 \kappa ^2 (\mu  {r_0})^{5/3}-332 \kappa  ({\mu  {r_0}})^{1/3}+36 \kappa  (\mu  {r_0})^{7/3}+48 \mu  {r_0}\right)
+ 3 {r_0}^2 \left(24 \kappa  (\mu  {r_0})^{7/3}-112 \kappa  ({\mu  {r_0}})^{1/3}\right.\nonumber\\
&&  \left.\left. + \kappa ^2 \mu ^3 {r_0}^3 \left(9 (\mu  {r_0})^{2/3}-70 \kappa \right)-4 \mu  {r_0} \left(45 \kappa ^2 (\mu  {r_0})^{2/3}-4\right)\right)\right]
\left[8 {M_0}^2 \left(161 \kappa  ({\mu  {r_0}})^{1/3}+24 \mu  {r_0}\right)-4 M_0 {r_0}\right.\nonumber\\
&& \times \left(273 \kappa ^2 (\mu {r_0})^{5/3}+332 \kappa  ({\mu  {r_0}})^{1/3}+36 \kappa  (\mu  {r_0})^{7/3}+48 \mu  {r_0}\right)
+3 {r_0}^2 \left(
\kappa ^2 \mu ^3 {r_0}^3 \left(70 \kappa +9 (\mu  {r_0})^{2/3}\right)
 \right. \nonumber\\
 && \left.\left. +112 \kappa  ({\mu  {r_0}})^{1/3}+24 \kappa  (\mu  {r_0})^{7/3}+4 \mu {r_0} \left(45 \kappa ^2 (\mu  {r_0})^{2/3}+4\right)\right)\right] + \left(1-{2 M_0}{r_0}^{-1}+\frac{3}{4} \kappa  (\mu  {r_0})^{4/3}\right)^{-1} \kappa ^3 \mu \left[ 644 {M_0}^2 ({\mu  {r_0}})^{1/3}\right.\nonumber\\
 && \left. -664 M_0 {r_0} ({\mu  {r_0}})^{1/3}+105 \kappa ^2 \mu ^3 {r_0}^5 -546 \kappa  M_0 {r_0} (\mu  {r_0})^{5/3}+270 \kappa  {r_0}^2 (\mu  {r_0})^{5/3}+168 {r_0}^2 ({\mu  {r_0}})^{1/3}\right]^5 \left\{1
   \right.\nonumber\\
 &&\left.
+ \left[ 
 9 \mu ^2 {r_0}^2 \left({r_0} \left(3 \kappa  (\mu  {r_0})^{4/3}+4\right)-8 M\right)^4\right] \left[ 4 \kappa ^2\left( 105 \kappa ^2 \mu ^3 {r_0}^5 
 +644 {M_0}^2 ({\mu  {r_0}})^{1/3}-546 \kappa  M_0 {r_0} (\mu  {r_0})^{5/3}-664 M_0 {r_0} ({\mu  {r_0}})^{1/3}\right.\right.\right.\nonumber\\
 &&\left.\left.\left. +270 \kappa  {r_0}^2 (\mu  {r_0})^{5/3}+168 {r_0}^2 ({\mu  {r_0}})^{1/3}\right)^2\right]^{-1}\right\} \geq 0.
\end{eqnarray}

\section{Derivation of the Generalized Raychaudhuri Equation for Non-Normalized Tangential Vectors}\label{app:DerivationRaychaudhuri}

We look for the evolution equation of $\bar B\indices{^\alpha_\beta}$, where
\begin{equation}\label{eq:barB-app}
\bar B\indices{^\alpha_\beta} = h\indices{_\mu^\alpha} h\indices{_\beta^\nu} 
B\indices{^\mu_\nu} = h\indices{_\mu^\alpha} h\indices{_\beta^\nu} u\indices{^\mu_{;\nu}}, 
\end{equation}
and $h\indices{_\beta^\alpha}$ has been defined in Eq.~(\ref{eq:halphabeta-NONnormalized-General}). The rate of change of $\bar B\indices{^\alpha_\beta}$ along the increment of $\tau$ (i.e. for a comoving observer who travels with the congruence), is then written as
\begin{eqnarray}\label{eq:dotbarB-1-app}
\frac{D \bar{B}\indices{^\alpha_\beta}}{\ed \tau} = \dot{\bar{B}}\indices{^\alpha_\beta} = \bar{B}\indices{^\alpha_{\beta;\sigma}} u^\sigma &=& \left(h\indices{_\mu^\alpha}h\indices{_\beta^\nu}u\indices{^\mu_{;\nu}}\right)_{;\sigma} u^\sigma\nonumber\\
&=&\left(h\indices{_\mu^\alpha_{;\sigma}}h\indices{_\beta^\nu} + h\indices{_\mu^\alpha}h\indices{_\beta^\nu_{;\sigma}}\right)u\indices{^\mu_{;\nu}}u^\sigma 
+h\indices{_\mu^\alpha}h\indices{_\beta^\nu}u\indices{^\mu_{;\nu;\sigma}}u^\sigma\nonumber\\
&=&\left(h\indices{_\mu^\alpha_{;\sigma}}h\indices{_\beta^\nu} + h\indices{_\mu^\alpha}h\indices{_\beta^\nu_{;\sigma}}\right)u\indices{^\mu_{;\nu}}u^\sigma 
+h\indices{_\mu^\alpha}h\indices{_\beta^\nu}\left(-R\indices{^\mu_{\lambda\nu\sigma}} u^\lambda+u\indices{^\mu_{;\sigma;\nu}}\right)u^\sigma\nonumber
\\
&=&\left(h\indices{_\mu^\alpha_{;\sigma}}h\indices{_\beta^\nu} + h\indices{_\mu^\alpha}h\indices{_\beta^\nu_{;\sigma}}\right)u\indices{^\mu_{;\nu}}u^\sigma +h\indices{_\mu^\alpha}h\indices{_\beta^\nu}
\left(-R\indices{^\mu_{\lambda\nu\sigma}} u^\lambda u^\alpha\right.
+ \left.\left(u\indices{^\mu_{;\sigma}} u^\sigma\right)_{;\nu} - u\indices{^\mu_{;\sigma}} u\indices{^\sigma_{;\nu}}\right)\nonumber\\
&=&\left(h\indices{_\mu^\alpha_{;\sigma}}h\indices{_\beta^\nu} + h\indices{_\mu^\alpha}h\indices{_\beta^\nu_{;\sigma}}\right)u\indices{^\mu_{;\nu}}u^\sigma + h\indices{_\mu^\alpha}h\indices{_\beta^\nu}a\indices{^\mu_{;\nu}}
-h\indices{_\mu^\alpha}h\indices{_\beta^\nu} R\indices{^\mu_{\lambda\nu\sigma}} u^\lambda u^\sigma
-\bar{B}\indices{^\alpha_\sigma}\bar{B}\indices{^\sigma_\beta},
\end{eqnarray}
in which to get from the second line to the third line, we have used $\left[\nabla_\nu,\nabla_\sigma\right] u^\mu = R\indices{^\mu_{\lambda\nu\sigma}} u^\lambda$. The first term of Eq.~(\ref{eq:dotbarB-1-app}), contains the two segments
\begin{subequations}
\begin{align}
h\indices{_\mu^\alpha_{;\sigma}} h\indices{_\beta^\nu} =\left(
\frac{-\left(u_{\mu;\sigma}u^\alpha+u_{\mu}u\indices{^\alpha_{;\sigma}}\right) u_\gamma u^\gamma + \left(u_{\gamma;\sigma}u^\gamma + u_\gamma u\indices{^\gamma_{;\sigma}}\right)u_\mu u^\alpha}{\left(u_\lambda u^\lambda\right)^2}\right)\left(\delta_\beta^\nu - \frac{u_\beta u^\nu}{u_\lambda u^\lambda}\right),
\label{eq:dotbarB-1-segment1}
\\
h\indices{_\mu^\alpha} h\indices{_\beta^\nu_{;\sigma}} =\left(\delta_\mu^\alpha - \frac{u_\mu u^\alpha}{u_\lambda u^\lambda}\right)\left(
\frac{-\left(u_{\beta;\sigma}u^\nu+u_{\beta}u\indices{^\nu_{;\sigma}}\right) u_\gamma u^\gamma + \left(u_{\gamma;\sigma}u^\gamma + u_\gamma u\indices{^\gamma_{;\sigma}}\right)u_\beta u^\nu}{\left(u_\lambda u^\lambda\right)^2}\right).\label{eq:dotbarB-1-segment2}
\end{align}
\end{subequations}
Therefore
\begin{eqnarray}\label{eq:dotbarB-1-segment1+2}
\left(h\indices{_\mu^\alpha_{;\sigma}}h\indices{_\beta^\nu} + h\indices{_\mu^\alpha}h\indices{_\beta^\nu_{;\sigma}}\right)u\indices{^\mu_{;\nu}}u^\sigma\nonumber\\
&&= \frac{\left(u_{\gamma;\sigma}u^\gamma + u_\gamma u\indices{^\gamma_{;\sigma}}\right)u_\mu u^\alpha - \left(u_{\mu;\sigma}u^\alpha+u_{\mu}u\indices{^\alpha_{;\sigma}}\right) u_\gamma u^\gamma}{\left(u_\lambda u^\lambda\right)^2}~ u\indices{^\mu_{;\beta}} u^\sigma\nonumber\\
&&-\frac{\left(u_{\gamma;\sigma}u^\gamma + u_\gamma u\indices{^\gamma_{;\sigma}}\right)u_\mu u^\alpha u_\beta u^\nu- \left(u_{\mu;\sigma}u^\alpha+u_{\mu}u\indices{^\alpha_{;\sigma}}\right) u_\gamma u^\gamma u_\beta u^\nu}{\left(u_\lambda u^\lambda\right)^3}~ u\indices{^\mu_{;\nu}} u^\sigma\nonumber\\
&& + \frac{\left(u_{\gamma;\sigma}u^\gamma + u_\gamma u\indices{^\gamma_{;\sigma}}\right)u_\beta u^\nu - \left(u_{\beta;\sigma}u^\nu+u_{\beta}u\indices{^\nu_{;\sigma}}\right) u_\gamma u^\gamma}{\left(u_\lambda u^\lambda\right)^2}~ u\indices{^\alpha_{;\nu}} u^\sigma\nonumber\\
&& - \frac{\left(u_{\gamma;\sigma}u^\gamma + u_\gamma u\indices{^\gamma_{;\sigma}}\right)u_\beta u^\nu u_\mu u^\alpha- \left(u_{\beta;\sigma}u^\nu+u_{\beta}u\indices{^\nu_{;\sigma}}\right) u_\gamma u^\gamma u_\mu u^\alpha}{\left(u_\lambda u^\lambda\right)^3}~ u\indices{^\mu_{;\nu}} u^\sigma.
\end{eqnarray}
In the third line of the above expressions, we change the dummy indices $\nu$ to $\mu$. Also considering $u\indices{^\alpha_{;\sigma}} u^\sigma= a^\alpha$, then Eq.~(\ref{eq:dotbarB-1-segment1+2}) changes to 
\begin{eqnarray}
\left(h\indices{_\mu^\alpha_{;\sigma}}h\indices{_\beta^\nu} + h\indices{_\mu^\alpha}h\indices{_\beta^\nu_{;\sigma}}\right)u\indices{^\mu_{;\nu}}u^\sigma\nonumber\\
&&= \frac{\left(a_{\gamma}u^\gamma + u_\gamma a^\gamma\right)u_\mu u^\alpha - \left(a_{\mu}u^\alpha+u_{\mu}a^\alpha\right) u_\gamma u^\gamma}{\left(u_\lambda u^\lambda\right)^2}~ u\indices{^\mu_{;\beta}} \nonumber\\
&&-\frac{\left(a_{\gamma}u^\gamma + u_\gamma a
^\gamma\right)u_\mu u^\alpha u_\beta u^\nu- \left(a_{\mu}u^\alpha+u_{\mu}a{^\alpha}\right) u_\gamma u^\gamma u_\beta u^\nu}{\left(u_\lambda u^\lambda\right)^3}~ u\indices{^\mu_{;\nu}} \nonumber\\
&& + \frac{\left(a_{\gamma}u^\gamma + u_\gamma a\indices{^\gamma}\right)u_\beta u^\mu - \left(a_{\beta}u^\mu+u_{\beta}a{^\mu}\right) u_\gamma u^\gamma}{\left(u_\lambda u^\lambda\right)^2}~ u\indices{^\alpha_{;\mu}} \nonumber\\
&& - \frac{\left(a_{\gamma}u^\gamma + u_\gamma a{^\gamma}\right)u_\beta u^\nu u_\mu u^\alpha- \left(a_{\beta}u^\nu+u_{\beta}a{^\nu}\right) u_\gamma u^\gamma u_\mu u^\alpha}{\left(u_\lambda u^\lambda\right)^3}~ u\indices{^\mu_{;\nu}}\nonumber\\\nonumber\\\nonumber\\
&& = \left[\frac{a_\gamma u^\gamma + u_\gamma a^\gamma}{\left(u_\lambda u^\lambda\right)^2}\left(u_\mu u^\alpha u\indices{^\mu_{;\beta}} + u^\mu u_\beta u\indices{^\alpha_{;\mu}} \right) - \frac{1}{u_\lambda u^\lambda}\left\{ \left(a_\mu u^\alpha + u_\mu a^\alpha\right) u\indices{^\mu_{;\beta}} \right.\right.\nonumber\\
&{}&\quad\left.\left. +  \left(a_\beta u^\mu + u_\beta a^\mu\right) u\indices{^\alpha_{;\mu}}
\right\}
\right]\nonumber\\
&& - \left[\frac{2\left(a_\gamma u^\gamma + u_\gamma a^\gamma\right) u^\alpha u_\beta u^\nu u_\mu}{\left(u_\lambda u^\lambda\right)^3}~u\indices{^\mu_{;\nu}}
-\frac{1}{\left(u_\lambda u^\lambda\right)^2} \left\{\left(a_\mu u^\alpha + u_\mu a^\alpha\right) u_\beta u^\nu u\indices{^\mu_{;\nu}}
\right.
\right.\nonumber\\
&&\quad \left.\left. + 
\left(a_\beta u^\nu + u_\beta a^\nu\right) u_\mu u^\alpha u\indices{^\mu_{;\nu}}
\right\}\right]\nonumber
\end{eqnarray}
\begin{eqnarray}\label{eq:dotbarB-2-segment1+2}
\quad\quad \quad\quad\quad\quad \quad\quad \quad\quad\quad\quad \quad\quad\quad & = &
\left[\frac{a_\gamma u^\gamma + u_\gamma a^\gamma}{\left(u_\lambda u^\lambda\right)^2} \left(\frac{1}{2}\left(u^\mu u_\mu\right)_{;\beta} u^\alpha + u_\beta a^\alpha\right)
-\frac{1}{u_\lambda u^\lambda} \left\{
a_\mu u\indices{^\mu_{;\beta}} u^\alpha + \frac{1}{2}\left(u^\mu u_\mu\right)_{;\beta} a^\alpha
\right.
\right.\nonumber\\
&&\left.\left.+ a_\beta a^\alpha + u_\beta a^\mu u\indices{^\alpha_{;\mu}}
\right\}
\right]\nonumber\\ 
&& - \left[
\frac{2 \left(a_\gamma u^\gamma + u_\gamma a^\gamma\right) u^\alpha u_\beta a^\mu u_\mu}{\left(u_\lambda u^\lambda\right)^3} - \frac{1}{\left(u_\lambda u^\lambda\right)^2}\left\{
\left(a_\mu u^\alpha + u_\mu a^\alpha\right) u_\beta a^\mu
\right.
\right.\nonumber\\
&&\left.\left.\quad +\frac{1}{2}\left(a_\beta u^\mu + u_\beta a^\mu\right) u^\alpha \left(u^\nu u_\nu\right)_{;\mu}
\right\}
\right]\nonumber\\\nonumber\\
 && = \frac{a_\gamma u^\gamma + u_\gamma a^\gamma}{\left(u_\lambda u^\lambda\right)^2} \left(\frac{1}{2} u^\alpha \left(u^\mu u_\mu\right)_{;\beta}\right)
+\frac{a_\gamma u^\gamma + u_\gamma a^\gamma}{\left(u_\lambda u^\lambda\right)^2}~u_\beta a^\alpha\nonumber\\
&& - \frac{1}{u_\lambda u^\lambda} \left\{
\left(u^\alpha a_\mu u\indices{^\mu_{;\beta}} + u_\beta a^\mu u\indices{^\alpha_{;\mu}}\right) + a^\alpha a_\beta + \frac{1}{2} \left(u^\mu u_\mu\right)_{;\beta} a^\alpha\right\}\nonumber\\
&& - \frac{2 \left(a_\gamma u^\gamma + u_\gamma a^\gamma\right) a^\mu u_\mu}{\left(u_\lambda u^\lambda\right)^3}~u^\alpha u_\beta 
+ \frac{a_\mu u^\alpha + u_\mu a^\alpha}{\left(u_\lambda u^\lambda\right)^2}~u_\beta a^\mu\nonumber\\
&&+\frac{1}{2} ~\frac{a_\beta u^\mu + u_\beta a^\mu}{\left(u_\lambda u^\lambda\right)^2}~u^\alpha \left(u^\nu u_\nu\right)_{;\mu}\nonumber\\\nonumber\\\nonumber\\
&&= \frac{u^\alpha}{2 \left(u_\lambda u^\lambda\right)^2}
\left[\left(a_\gamma u^\gamma + u_\gamma a^\gamma\right)
\left(u^\mu u_\mu\right)_{;\beta}
+\left(a_\beta u^\mu + u_\beta a^\mu\right)\left(u^\nu u_\nu\right)_{;\mu}
\right]\nonumber\\
&&+ \frac{u_\beta}{\left(u_\lambda u^\lambda\right)^2}
\left[
\left(a_\gamma u^\gamma + u_\gamma a^\gamma\right) a^\alpha
+\left(a_\mu u^\alpha + u_\mu a^\alpha\right) a^\mu
\right]\nonumber\\
&& - \frac{1}{u_\lambda u^\lambda} \left[
\frac{1}{2} \left(u^\mu u_\mu\right)_{;\beta} a^\alpha
+a^\alpha a_\beta + u^\alpha a_\mu u\indices{^\mu_{;\beta}} + u_\beta a^\mu u\indices{^\alpha_{;\mu}}
\right]\nonumber\\
&&-\frac{2\left(a_\gamma u^\gamma + u_\gamma a^\gamma\right) a^\mu u_\mu}{\left(u_\lambda u^\lambda\right)^3}~u^\alpha u_\beta.
\end{eqnarray}
Substitution of Eq.~(\ref{eq:dotbarB-2-segment1+2}) in Eq.~(\ref{eq:dotbarB-1-app}) gives
\begin{eqnarray}\label{eq:dotbarB-2-app}
\dot{\bar{B}}\indices{^\alpha_\beta} & = & \frac{u^\alpha}{2 \left(u_\lambda u^\lambda\right)^2}
\left[2 a_\gamma u^\gamma 
\left(u^\mu u_\mu\right)_{;\beta}
+\left(a_\beta u^\mu + u_\beta a^\mu\right)\left(u^\nu u_\nu\right)_{;\mu}
\right]\nonumber\\
&&+ \frac{u_\beta}{\left(u_\lambda u^\lambda\right)^2}
\left[
2 a_\gamma u^\gamma  a^\alpha
+\left(a_\mu u^\alpha + u_\mu a^\alpha\right) a^\mu
\right]\nonumber\\
&& - \frac{1}{u_\lambda u^\lambda} \left[
\frac{1}{2} \left(u^\mu u_\mu\right)_{;\beta} a^\alpha
+a^\alpha a_\beta + u^\alpha a_\mu u\indices{^\mu_{;\beta}} + u_\beta a^\mu u\indices{^\alpha_{;\mu}}
\right]\nonumber\\
&&-\frac{\left(2 a_\gamma u^\gamma \right)^2}{\left(u_\lambda u^\lambda\right)^3}~u^\alpha u_\beta  + h\indices{_\mu^\alpha}h\indices{_\beta^\nu}a\indices{^\mu_{;\nu}}
-h\indices{_\mu^\alpha}h\indices{_\beta^\nu} R\indices{^\mu_{\lambda\nu\sigma}} u^\lambda u^\sigma
-\bar{B}\indices{^\alpha_\sigma}\bar{B}\indices{^\sigma_\beta}.
\end{eqnarray}
The evolution of $\Theta$ along the congruence, is then given by the trace of the above object. Hence, we have
\begin{eqnarray}\label{eq:dotbarB-3-app}
\dot{\bar{B}}\indices{^\alpha_\alpha}  \equiv \dot\Theta & = & \frac{u^\alpha}{2 \left(u_\lambda u^\lambda\right)^2}
\left[2 a_\gamma u^\gamma 
\left(u^\mu u_\mu\right)_{;\alpha}
+\left(a_\alpha u^\mu + u_\alpha a^\mu\right)\left(u^\nu u_\nu\right)_{;\mu}
\right]\nonumber\\
&&+ \frac{u_\alpha}{\left(u_\lambda u^\lambda\right)^2}
\left[
2 a_\gamma u^\gamma  a^\alpha
+\left(a_\mu u^\alpha + u_\mu a^\alpha\right) a^\mu
\right]\nonumber\\
&& - \frac{1}{u_\lambda u^\lambda} \left[
\frac{1}{2} \left(u^\mu u_\mu\right)_{;\alpha} a^\alpha
+a^\alpha a_\alpha + u^\alpha a_\mu u\indices{^\mu_{;\alpha}} + u_\alpha a^\mu u\indices{^\alpha_{;\mu}}
\right]\nonumber\\
&&-\left(\frac{2 a_\gamma u^\gamma}{u_\lambda u^\lambda}\right)^2 + h\indices{_\mu^\alpha}h\indices{_\alpha^\nu}a\indices{^\mu_{;\nu}}
-h\indices{_\mu^\alpha}h\indices{_\alpha^\nu} R\indices{^\mu_{\lambda\nu\sigma}} u^\lambda u^\sigma
-\bar{B}\indices{^\alpha_\sigma}\bar{B}\indices{^\sigma_\alpha}.
\end{eqnarray}
It is straightforward to justify that for the tensor $\bar{B}\indices{^\alpha_\beta} = 1/3 ~\Theta h\indices{_\beta^\alpha} + \sigma\indices{^\alpha_\beta} + \omega\indices{^\alpha_\beta} $, we have
\begin{eqnarray}\label{eq:barB*barB-app}
\bar{B}\indices{^\alpha_\sigma}\bar{B}\indices{^\sigma_\alpha} &=&\frac{1}{3} \Theta^2 + \sigma\indices{^\alpha_\sigma} \sigma\indices{^\sigma_\alpha}+ \omega\indices{^\alpha_\sigma} \omega\indices{^\sigma_\alpha}\nonumber\\
& =& \frac{1}{3} \Theta^2 + \sigma_{\alpha\sigma} \sigma^{\alpha\sigma}
-\omega_{\alpha\sigma} \omega^{\alpha\sigma}.
\end{eqnarray}
Applying this, together with the identity $h\indices{_\mu^\alpha} h\indices{_\alpha^\nu} = h\indices{_\mu^\nu}$, we arrive at the Raychaudhuri equation:
\begin{eqnarray}\label{eq:RaychaudhuriEqn-app}
\dot \Theta  & = & -\frac{1}{3} \Theta^2 -\sigma_{\alpha\beta} \sigma^{\alpha\beta} + \omega_{\alpha\beta} \omega^{\alpha\beta} + h\indices{_\beta^\alpha} a\indices{^\beta_{;\alpha}}
-h\indices{_\beta^\alpha} R\indices{^\beta_{\mu\alpha\nu}} u^\mu u^\nu\nonumber\\
&& + \frac{u^\alpha}{2(u_\lambda u^\lambda)^2}\left[2 a_\gamma u^\gamma \left(2 a_\alpha + (u^\mu u_\mu)_{;\alpha}\right) + 
\left(a_\alpha u^\mu + u_\alpha a^\mu\right) \left(2 a_\mu + (u^\nu u_\nu)_{;\mu}\right)
\right]\nonumber\\
&&-\frac{1}{u_\lambda u^\lambda} \left[2 a^\mu a_\mu + (u^\mu u_\mu)_{;\alpha} a^\alpha \right] - \left(\frac{2  a_\gamma u^\gamma}{u_\lambda u^\lambda}\right)^2.
\end{eqnarray}
To calculate the scalar expansion $\Theta$, let us expand $\bar{B}\indices{^\alpha_\beta}$ in Eq.~(\ref{eq:barB-app}). We have
\begin{eqnarray}\label{eq:barB-2-app}
\bar{B}\indices{^\alpha_\beta} &=&
\left(\delta_\mu^\alpha \delta_\beta^\nu
-\frac{u_\beta u^\nu \delta_\mu^\alpha}{u_\lambda u^\lambda}
-\frac{u_\mu u^\alpha \delta_\beta^\nu}{u_\lambda u^\lambda}
+\frac{u_\mu u^\nu u^\alpha u_\beta}{\left(u_\lambda u^\lambda\right)^2} 
\right) u\indices{^\mu_{;\nu}}\nonumber\\
& =& u\indices{^\alpha_{;\beta}} - \frac{u_\beta a^\alpha}{u_\lambda u^\lambda}
-\frac{\left(u^\mu u_\mu\right)_{;\beta} u^\alpha}{2 u_\lambda u^\lambda}
+\frac{u^\alpha u_\beta a^\mu u_\mu}{\left(u_\lambda u^\lambda\right)^2},
\end{eqnarray}
which satisfies $u_\alpha \bar{B}\indices{^\alpha_\beta} = \bar{B}\indices{^\alpha_\beta} u^\beta =0$. The scalar expansion is the trace of the above relation.
\begin{eqnarray}\label{eq:barB-3-app}
\Theta \equiv \bar{B}\indices{^\alpha_\alpha} = u\indices{^\alpha_{;\alpha}} - \frac{u^\alpha a_\alpha}{ u_\lambda u^\lambda}.
\end{eqnarray}	
For any congruence whose tangential vector satisfies $u^\alpha u_\alpha = \mathrm{const.}$, then $u^\alpha a_\alpha  = u^\alpha u_{\alpha;\lambda} u^\lambda = 1/2~\left(u^\alpha u_\alpha\right)_{;\lambda} u^\lambda =0$. If we define $\widetilde{h}\indices{_\beta^\alpha}$, $\widetilde{\Theta}$, $\widetilde{\sigma}_{\alpha\beta}$ and $\widetilde{\omega}_{\alpha\beta}$ to denote the projection operator and the kinematical characteristics of such congruences, then the Raychaudhuri equation (\ref{eq:RaychaudhuriEqn-app}) reduces to 
\begin{eqnarray}\label{eq:RaychaudhuriEqn-app-reduced-1}
\dot{\widetilde{\Theta}} &=& -\frac{1}{3}\widetilde{\Theta}^2 - \widetilde{\sigma}_{\alpha\beta} \widetilde{\sigma}^{\alpha\beta}
+\widetilde{\omega}_{\alpha\beta} \widetilde{\omega}^{\alpha\beta}
+ \widetilde{h}\indices{_\beta^\alpha} a\indices{^\beta_{;\alpha}}-
\widetilde{h}\indices{_\beta^\alpha} R\indices{^\beta_{\mu\alpha\nu}} u^\mu u^\nu\nonumber\\
&&+\frac{u^\alpha}{2} \left(2 u_\alpha a^\mu a_\mu\right) + 2 a^\mu a_\mu \nonumber\\\nonumber\\
& =& -\frac{1}{3}\widetilde{\Theta}^2 - \widetilde{\sigma}_{\alpha\beta} \widetilde{\sigma}^{\alpha\beta}
+\widetilde{\omega}_{\alpha\beta} \widetilde{\omega}^{\alpha\beta}
+  a\indices{^\alpha_{;\alpha}} 
+ u_\beta u^\alpha \left(u\indices{^\beta_{;\lambda}} u^\lambda\right)_{;\alpha}\nonumber\\
&& - R_{\mu\nu} u^\mu u^\nu - R\indices{^\beta_{\mu\alpha\nu}} u_\beta u^\mu u^\alpha u^\nu + a^\mu a_\mu,
\end{eqnarray}
where $\widetilde\Theta = u\indices{^\alpha_{;\alpha}}$. By virtue of the anti-symmetries of the Riemann tensor, we have $R\indices{^\beta_{\mu\alpha\nu}} u_\beta u^\mu u^\alpha u^\nu = 0$. Furthermore, we can do the following simplification:
\begin{eqnarray}\label{eq:TheLastTerm-3rdLine-Simplified-app}
u_\beta u^\alpha \left(u\indices{^\beta_{;\lambda}} u^\lambda\right)_{;\alpha}
&=& u_\beta u^\alpha \left(u\indices{^\beta_{;\lambda;\alpha}} u^\lambda + u\indices{^\beta_{;\lambda}} u\indices{^\lambda_{;\alpha}}\right)\nonumber\\
& = & u^\lambda u^\alpha \left[\left(u\indices{^\beta_{;\lambda}}u_\beta\right)_{;\alpha} - u\indices{^\beta_{;\lambda}} u_{\beta;\alpha}\right] 
+ \frac{1}{2} \left(u^\beta u_\beta\right)_{;\lambda} a^\lambda\nonumber\\
& = & u^\lambda u^\alpha \left[\left(\frac{1}{2}\left(u^\beta u_\beta\right)_{;\lambda}\right)_{;\alpha} - u\indices{^\beta_{;\lambda}} u_{\beta;\alpha}\right]\nonumber\\
& = & -a^\beta a_\beta.
\end{eqnarray}
Interpolation in Eq.~(\ref{eq:RaychaudhuriEqn-app-reduced-1}) results in
\begin{eqnarray}\label{eq:RaychaudhuriEqn-app-reduced-2}
\dot{\widetilde{\Theta}} 
= -\frac{1}{3}\widetilde{\Theta}^2 - \widetilde{\sigma}_{\alpha\beta} \widetilde{\sigma}^{\alpha\beta}
+\widetilde{\omega}_{\alpha\beta} \widetilde{\omega}^{\alpha\beta}
- R_{\alpha\beta} u^\alpha u^\beta 
+ a\indices{^\alpha_{;\alpha}},
\end{eqnarray}
which is the usual form of the Raychaudhuri equation for normalized tangential vectors, as it is in Eq.~(\ref{eq:Raychaudhuri-Common}).

\section{The Algebraic Expression of the SEC for the Non-Geodesic Congruence}

\begin{eqnarray}\label{eq:SEC-nongeodesic-app}
\mathcal{M} &\doteq& 
\frac{9 \kappa  \mu ^3 {r_s}^4 \left({r_s} \left(3 \kappa  (\mu  {r_s})^{4/3}+4\right)-8 M_s\right)^4}{644 {M_s}^2 \left(\mu r_s\right)^{1/3}-546 \kappa  M_s {r_s} (\mu  {r_s})^{5/3}-664 M_s {r_s} \left(\mu r_s\right)^{1/3}+105 \kappa ^2 \mu ^3 {r_s}^5+270 \kappa  {r_s}^2 (\mu  {r_s})^{5/3}+168 {r_s}^2 \left(\mu r_s\right)^{1/3}}\nonumber\\
&& + \mathcal{F}^{-2} \left[
\kappa  {r_s} (\mu  {r_s})^{2/3}\left\{ -64 {M_s}^4
\left(576 \mu ^2 {r_s}^2-25921 \kappa ^2 (\mu  {r_s})^{2/3}\right)
+64 {M_s}^3 \left(
3 {r_s}^3 \left(384-14651 \kappa ^3\right) \mu ^2
+864 \kappa  \mu ^3 {r_s}^4 \left(\mu r_s\right)^{1/3}
\right.
\right.
\right.\nonumber\\
&&\left.\left.\left. -53452 \kappa ^2 {r_s} (\mu  {r_s})^{2/3}\right)
- 16 {M_s}^2 \left(12 {r_s}^4
\left(288-22351 \kappa ^3\right) \mu ^2+3 \kappa  \left(1728-36113 \kappa ^3\right) \mu ^3 {r_s}^5 \left(\mu r_s\right)^{1/3}
-164320 \kappa ^2 {r_s}^2 (\mu {r_s})^{2/3}
\right.
\right.
\right.\nonumber\\
&&\left.\left.\left.
+1944 \kappa ^2 \mu ^4 {r_s}^6 (\mu  {r_s})^{2/3}
\right)
+ 48 M_s \left(
162 \kappa ^3 \mu ^6 {r_s}^9+2 \kappa  \left(432-18095 \kappa ^3\right) \mu ^3 {r_s}^6 \left(\mu r_s\right)^{1/3}-48 \left(941 \kappa ^3-8\right) \mu ^2 {r_s}^5
\right.
\right.
\right.\nonumber\\
&&\left.\left.\left.
-18592 \kappa ^2 {r_s}^3 (\mu  {r_s})^{2/3}+3 \kappa ^2 \left(216-3185 \kappa ^3\right) \mu ^4 {r_s}^7 (\mu  {r_s})^{2/3}
\right)
- 9 {r_s}^4 \left(
4 \kappa ^3 \left(108-1225 \kappa ^3\right) \mu ^6 {r_s}^6
\right.
\right.
\right.\nonumber\\
&&\left.\left.
+16 \kappa  \left(48-3005 \kappa ^3\right) \mu ^3 {r_s}^3 \left(\mu r_s\right)^{1/3}-128 \left(315 \kappa ^3-2\right) \mu ^2 {r_s}^2+81 \kappa ^4 \mu ^7 {r_s}^7 \left(\mu r_s\right)^{1/3}-12544 \kappa ^2 (\mu  {r_s})^{2/3}
\right.
\right.\nonumber\\
&&\left.\left.\left.
-144 \kappa ^2 \left(175 \kappa ^3-6\right) \mu ^4 {r_s}^4 (\mu  {r_s})^{2/3}\right)\right\}
\left(
154 {M_s}^2-2 M_s {r_s} \left(63 \kappa  (\mu  {r_s})^{4/3}+62\right)
+{r_s}^2 \left(21 \kappa ^2 (\mu  {r_s})^{8/3}+57 \kappa  (\mu  {r_s})^{4/3}+28\right)
\right)
\right]\nonumber\\
&&
 + \mathcal{F}^{-4} \left[
3 (\mu  {r_s})^{4/3} \left(-2 M_s+\kappa  {r_s} (\mu  {r_s})^{4/3}+{r_s}\right)\left({r_s} \left(3 \kappa  (\mu  {r_s})^{4/3}+4\right)-8 M_s\right)^3
\left({r_s} \left(9 \kappa  (\mu {r_s})^{4/3}+4\right)-24 M_s\right)
\right.\nonumber\\
&&\times 
\left\{
26543104 {M_s}^6 +41216 {M_s}^5 {r_s} \left(1071 \kappa  (\mu  {r_s})^{4/3}-1732\right)
+768 {M_s}^4 {r_s}^2 \left(
119116-40383 \kappa  (\mu  {r_s})^{4/3}
+10045 \kappa ^2 (\mu  {r_s})^{8/3}\right)
\right.\nonumber\\
&&\left.
+ 9 {r_s}^6
\left(
11025 \kappa ^6 \mu ^8 {r_s}^8+254700 \kappa ^4 \mu ^5 {r_s}^5 \left(\mu r_s\right)^{1/3}+372960 \kappa ^3 \mu ^4 {r_s}^4+50176
+84000 \kappa ^5 \mu ^6 {r_s}^6 (\mu  {r_s})^{2/3}
+286720 \kappa ^2 (\mu  {r_s})^{8/3}
\right.
\right.\nonumber\\
&&\left.\left.
+139776 \kappa  (\mu  {r_s})^{4/3}
\right)
- 6 M_s {r_s}^5 
\left(
275625 \kappa ^5 \mu ^6 {r_s}^6 (\mu  {r_s})^{2/3}+1729350 \kappa ^4 \mu ^5 {r_s}^5 \left(\mu r_s\right)^{1/3}+3607152 \kappa ^3 \mu ^4 {r_s}^4+885248
\right.
\right.\nonumber\\
&&\left.\left.
+3267712 \kappa ^2 (\mu  {r_s})^{8/3}+1784832 \kappa  (\mu  {r_s})^{4/3}
\right)
- 16 {M_s}^3 {r_s}^3
\left(
2135763 \kappa ^3 \mu ^4 {r_s}^4+3361176 \kappa ^2 (\mu  {r_s})^{8/3}+1668528 \kappa  (\mu  {r_s})^{4/3}
\right)
\right.\nonumber\\
&&\left.\left.
+4073216
+ 2 {M_s}^2 {r_s}^4
\left(
6687765 \kappa ^4 \mu ^5 {r_0}^5 \left(\mu r_s\right)^{1/3}+24637788 \kappa ^3 \mu ^4 {r_s}^4+16043328 \kappa  (\mu  {r_s})^{4/3}+12905984
\right.
\right.
\right.\nonumber\\
&&\left.\left.\left.
+27289872 \kappa ^2 (\mu  {r_s})^{8/3}
\right)
\right\}
\right]
 \geq 0,
\end{eqnarray}
in which 
$$\mathcal F = 644 {M_s}^2-2 M_s {r_s} \left(273 \kappa  (\mu  {r_s})^{4/3}+332\right)+3 {r_s}^2 \left(35 \kappa ^2 (\mu  {r_s})^{8/3}+90 \kappa  (\mu  {r_s})^{4/3}+56\right).$$


%

\end{document}